\documentclass[a4paper,11pt]{article}
\pdfoutput=1 

\usepackage{jheppub} 

\usepackage[T1]{fontenc} 
\usepackage{units}
\usepackage{subcaption}
\usepackage{float}
\RequirePackage[dvipsnames]{xcolor}

\newcommand{\jewel}{\textsc{Jewel}}

\title{\boldmath Towards an unbiased jet energy loss measurement}

\author[a,b]{Liliana Apolinário,}
\author[a,b]{Lénea Luís,}
\author[a,b]{José Guilherme Milhano,}
\author[a,b]{João M. Silva}

\affiliation[a]{LIP, Av. Prof. Gama Pinto, 2, P-1649-003 Lisboa, Portugal}
\affiliation[b]{Instituto Superior Técnico (IST), Universidade de Lisboa,  Av. Rovisco Pais 1, P-1049-001 Lisboa, Portugal}

\emailAdd{liliana@lip.pt}
\emailAdd{lenealuis@tecnico.ulisboa.pt}
\emailAdd{joao.m.da.silva@tecnico.ulisboa.pt}
\emailAdd{gmilhano@lip.pt}

\abstract{The modifications imprinted on jets due to their interaction with Quark Gluon Plasma (QGP) are assessed by comparing samples of jets produced in nucleus-nucleus collisions and proton-proton collisions. 
The standard procedure ignores the effect of bin migration by comparing specific observables for jet populations at the same reconstructed jet transverse momentum ($p_T$). 
Since jet $p_T$ is itself modified by interaction with QGP, all such comparisons confound QGP induced modifications with changes that are simply a consequence of comparing jets that started out differently. 
The quantile matching procedure introduced by Brewer et al. directly estimates average fractional jet energy loss ($Q_{AA}$) and can thus mitigate this $p_T$ migration effect. 
In this work, we validate the procedure in more realistic scenarios that include medium response. 
We study the evolution of $Q_{AA}$ with jet radius, its sensitivity to minimum particle $p_T$ and medium response as implemented in two different models for jet evolution in heavy-ion collisions. 
Further, we use this procedure to establish that the difference between inclusive jet and $\gamma+$jet nuclear modification factors ($R_{AA}$) is dominated by differences in the spectral shape, leaving the colour charge of the jet initiating parton with a lesser role to play.
Additionally, we compare $Q_{AA}$ to an experimentally proposed proxy for fractional jet energy loss, $S_{loss}$, showing that both quantities are similar, although the former provides a more clear physical interpretation.
Finally, we show the size of the $p_T$ migration correction for four different substructure observables and how to reliably use the quantile procedure experimentally to improve existing measurements.
}

\begin{document} 
\maketitle
\flushbottom

\section{Introduction}
\label{sec:intro}

From the structure of Quantum Chromodynamics (QCD), the fundamental theory of strong interactions, a particularly interesting system emerges: the quark-gluon plasma (QGP), a hot and dense state of matter where quarks and gluons are not confined inside hadrons. This system is believed to have been produced in the early epochs of the Universe and, experimentally, it has been recreated in the aftermath of ultrarelativistic heavy ion collisions at the Relativistic Heavy Ion Collider (RHIC)~\cite{PHENIX:2001hpc, STAR:2005gfr, STAR:2020xiv} and the Large Hadron Collider (LHC) ~\cite{Muller:2012zq,ATLAS:2018gwx, CMS:2021vui}. 
These experimental programmes have allowed the exploration of the many-body properties of QCD in such extreme regimes, and led to the remarkable discovery that the QGP is a nearly ideal liquid. About 1 fm/c ($\sim 3$ ys) after a heavy ion collision, QGP production is signaled by an abrupt increase in both particle and energy densities, with energy densities well above those expected inside typical hadrons. After its production, the QGP acquires hydrodynamical behaviour and expands, losing its collective properties on a timescale of roughly 10 fm/c (see e.g.~\cite{Busza:2018rrf} for a comprehensive review of heavy-ion collisions).

To be able to quantitatively describe the properties of the QGP and their time evolution, a probe which is produced concurrently is needed. For this it is common to use highly energetic probes~\cite{Apolinario:2022vzg} produced in the rare high transverse momentum scatterings happening between the quark and gluon constituents of the nucleons in the nucleus. The most powerful ones are hadronic jets, the result of the fragmentation of highly boosted quarks and gluons. Jets are extended objects in space and time, since their fragmentation evolution develops through multiple energy scales from the initial hard scattering ($\gtrsim 0.1$ TeV) down to soft non-perturbative scales of hadronization ($ \lesssim 0.1$ GeV). For this reason, they are sensitive to the different stages of the QGP’s evolution. The vacuum description of jet evolution, as it happens in proton-proton (pp) collisions, is well established perturbatively through precise resummations of large logarithms in the form of parton showers, but jets in nuclei collisions are different because they interact with the produced background medium in a non-trivial way, stimulating processes which compete with the vacuum-like development of the ensuing cascade~\cite{Zapp:2012ak,Caucal:2018dla}. The imprints of these interactions are carried through to the final particle distribution inside jets, with part of the energy thermalizing to the medium. This implies that jets reconstructed in nucleus-nucleus (AA) collisions are quenched versions of pp jets plus additional correlated background stemming from the medium’s response to the jet’s passage. The study of these medium induced effects is referred to as jet quenching~\cite{Mehtar-Tani:2013pia, Qin:2015srf, Cunqueiro:2021wls, Apolinario:2022vzg,Apolinario:2024equ} and it provides the framework through which one can compare jets produced in pp collisions with jets in AA collisions with the final aim of inferring the properties of the QGP.  

Despite the simplicity of this reasoning, fully describing heavy-ion jets is not an easy task since it requires formulating a parton shower in the medium which encapsulates all the relevant medium-induced effects while providing an accurate evolution model for the background medium. Not only this, but before following through with this reasoning, a simple yet very relevant question should first be answered: how exactly should one compare samples of vacuum and medium jets? The standard procedure to draw this comparison consists on selecting a window of reconstructed jet transverse momentum ($p_T$) and calculating the value of a given observable for both populations of jets. However, if our aim is to quantify jet modifications with respect to a vacuum baseline and if those modifications depend either explicitly or implicitly on $p_T$, this approach is biased. The key problem is that it ignores that medium jets migrate to lower $p_T$, i.e., jets lose energy as a result of interaction with QGP. Thus, comparisons drawn with this procedure confound QGP induced modifications with changes that are simply a consequence of comparing jets that were originated by hard partons with different $p_T$. To perform an unbiased comparison of jet populations, one thus needs to have an estimate of jet energy loss. 

In order to quantify how much energy jets in AA collisions lose with respect to their pp counterparts, it is useful to understand the physical mechanisms contributing to this effect. Interactions between the jet and QGP can generally be understood as momentum exchanges of the jet cascade with the medium. Among other effects, these exchanges result in elastic energy loss~\cite{Wicks:2005gt}, part of which thermalizes to the medium. The deposited energy, being correlated with the jet’s direction of propagation, can end up being reconstructed along with the final state jet constituents. One usually refers to this contribution to a heavy-ion jet as medium response. This way, depending on the jet reconstruction radius, part of the lost energy in elastic processes is recovered. 
The accumulation of successive scatterings of the partons in the evolving shower with the QGP can resolve fluctuations in their wavefunction into real radiation.
This medium-induced radiation has been tracted analytically within several perturbative schemes~\cite{Gyulassy:2000fs, Gyulassy:2000er},~\cite{Baier:1996kr, Baier:1998yf, Baier:1998kq, Zakharov:1996fv, Zakharov:1997uu},~\cite{Wang:2001ifa, Majumder:2009zu} and~\cite{Arnold:2001ms,Arnold:2002ja}. Regardless of the treatment, successive emissions stimulated by the medium are typically not angular ordered like what is predicted in a vacuum cascade~\cite{Mehtar-Tani:2010ebp}. Hence, part of these emissions have large angles with respect to the jet, falling outside of the reconstruction radius and significantly changing the jet energy distribution.

We emphasize that in this work we study jet energy loss and not parton energy loss and that this quantity has an important interplay with the jet radius~\cite{Mehtar-Tani:2021fud}. This means that the energy lost by jets reconstructed in AA collisions is with respect to the baseline of jets reconstructed in pp collisions, the energy of which is already a fraction of the initiating parton’s energy due to part of the vacuum shower falling outside of the jet reconstruction radius \cite{CDF:2005prv}. Recent works~\cite{He:2018gks, Wu:2023azi, Zhang:2023oid} have tried to quantify jet energy loss as well as correlate it with path length and partonic flavour in a model-independent way, by parameterizing an energy loss distribution and extracting the relevant parameters through Bayesian inference.

Experimentally, the nuclear modification factor $R_{AA}$, which is useful to quantify jet suppression, is often also used to qualitatively describe jet energy loss. It cannot, however, directly quantify jet energy loss, given that it takes the ratio of AA over pp cross sections at the same reconstructed $p_T$, ignoring $p_T$ migration. This bias is exacerbated by the steeply falling nature of the spectrum, to which the $R_{AA}$ is sensitive. In fact, if we consider two jet spectra with significantly different steepnesses, e.g. inclusive jet and boson+jet samples, and impose a fixed energy loss for all jets, then the less steeper spectrum will present a smaller suppression, i.e., $R_{AA}$ closer to 1. Hence, for the \textit{same} energy loss one obtains \textit{different} $R_{AA}$ because the shape of the spectrum was varied. Another problem caused by $p_T$ migration is selection bias (see e.g., \cite{Connors:2017ptx}), whereby samples of in-medium jets selected in a given $p_T$ window are biased towards those jets that were less modified due to the spectrum steepness. Both these problems can be partially evaded in boson+jet events~\cite{Brewer:2021hmh}, because there we have a reference for the $p_T$ of the hard parton that originated the jet. This way, one can compare pp and AA jets that started out with the same $p_T$. However, the size of the statistical sample of boson+jet events is significantly smaller than that of inclusive jet events, which one would like to use\footnote{Recently, in \cite{Andres:2024hdd} the problem of $p_T$ migration bias has been tackled in the context of energy correlator-based observables for inclusive jet events.}. We therefore need another solution. 
~\\

This paper is organized as follows. In Section~\ref{sec:qaa_intro} we introduce the quantile procedure as it was defined in the original paper~\cite{Brewer:2018dfs}, further motivating it as a way to calculate an energy loss proxy and clarifying the ideal scenarios where it works exactly. We also relate it to other existing energy loss proxies. Then, in Section~\ref{sec:generation}, we detail the physics implemented on the generators used to obtained the event samples analysed in this work, as well as the the relevant parameters for both the event generation setup and the analysis underlying jet reconstruction. In Section~\ref{sec:qaa_validation}, the quantile procedure is validated in a more realistic scenario than in~\cite{Brewer:2018dfs}, i.e., including medium response and for multiple jet radii. The following results in Section~\ref{sec:results} consist on studying, for two different event generators, how the energy loss proxy obtained via quantile matching varies with jet radius and with minimum particle $p_T$ cut. Additionally, a study of the colour charge dependence of energy loss is carried out, focusing on disentangling the spectrum steepness influence (which is large in $R_{AA}$) from the actual energy lost by jets. This is done for $\gamma+$jet, $\gamma+q$, $\gamma+g$ and inclusive jet samples. In Section~\ref{sec:measurement}, an obvious obstacle to the experimental implementation of the quantile procedure, namely the $p_T$ cutoff of the spectrum, is addressed and solved. A comparison with another energy loss proxy used in~\cite{ATLAS:2023iad} is made. Section~\ref{sec:substructure} shows the degree to which some jet substructure observables are changed when applying the quantile procedure to reduce the $p_T$ migration bias. Conclusions are presented in Section~\ref{sec:conclusions}.

\section{Quantile matching}\label{sec:qaa_intro}

Quantile matching as a means to estimate the average energy lost by jets through interaction with a QGP was introduced in \cite{Brewer:2018dfs}. The probability of a jet having transverse momentum within a given $(p_{T,1}, p_{T,2})$ interval is given by the cumulative jet cross section 
\begin{equation}
    \Sigma(p_{T,1}, p_{T,2}) = \int_{p_{T,1}}^{p_{T,2}}dp_T \frac{d\sigma}{dp_T}\, .
\end{equation}
The matching procedure aims at establishing a correspondence between a given transverse momentum ($p_T^v$) in the pp jet spectrum and a transverse momentum ($p_T^{q}$) in the AA jet spectrum by equating appropriately normalized upper cumulative cross sections
\begin{equation}\label{eq:quantile_matching}
    \Sigma^{pp}(p_T^v, +\infty) = \Sigma^{AA}(p_T^{q}(p_T^v), +\infty)\, ,
\end{equation}
i.e., by calculating a quantile function $p_T^{q}$ which depends on the value of $p_T^v$. 
This translates to finding out $p_T^{q}$ for a given $p_T^v$ such that AA jets with $p_T > p_T^{q} (p_T^v)$ are produced with the same probability as pp jets with $p_T > p_T^v$. 
As argued in the original paper \cite{Brewer:2018dfs}, in a scenario where a jet with a given $p_T$ always loses the same energy $\epsilon(p_T)$ and where that loss of energy either decreases or grows sufficiently slow with $p_T$ (i.e., $d\epsilon \left/\right. dp_T < 1$), the difference $p_T^v-p_T^q(p_T^v)$ is equal to $\epsilon(p_T^v)$. 
This is simply a consequence of the conservation of number of jets in the case where jet transverse momenta are exactly and identically ordered in pp and AA. 
In reality, the energy lost by jets originating from hard partons with the same $p_T$ can vary considerably \cite{Milhano:2015mng, Rajagopal:2016uip, Brewer:2017fqy, Casalderrey-Solana:2016jvj}.
The probabilistic nature of QCD branching and of the interaction of partons with the QGP, together with the varying amount of traversed QGP, all contribute to disperse the energy loss around some mean value \cite{He:2018gks} and away from the ideally ordered scenario. Thus, in \cite{Brewer:2018dfs} the authors explored the accuracy of
\begin{align}
    1-Q_{AA}(p_T^v) = 1-\frac{p_T^q(p_T^v)}{p_T^v} \, ,
\end{align}
as a proxy for the \emph{average} fractional jet energy loss of jets.

A crude analytic estimate of $Q_{AA}$ can be obtained as follows.\footnote{It is not within the scope of this paper to present detailed analytical calculations of $Q_{AA}$. Energy loss estimates using specific models were obtained in \cite{Takacs:2021bpv}.}
Assume that the jet spectrum in pp is described by a single power law
\begin{equation}\label{eq:assumption1}
    \frac{d\sigma^{pp}}{dp_T}(p_T) = A\, p_T^{-n}\, ,
\end{equation}
where $A$ is a normalization constant, and that the quenched jet spectrum can be obtained by taking an energy loss $\epsilon$ for each jet in the pp spectrum with probability $P(\epsilon, p_T)$ \cite{Baier:2001yt}:
\begin{equation}\label{eq:assumption2}
    \frac{d\sigma^{AA}}{dp_T}(p_T) = \int_0^{+\infty}d\epsilon \, P(\epsilon,p_T)\frac{d\sigma^{pp}}{dp_T}(p_T+\epsilon)
    \approx A\left(p_T + \langle \epsilon \rangle (p_T)\right)^{-n}\, .
\end{equation}
In the last equality, we assumed that the fractional energy loss $\epsilon \left /\right. p_T$ is small, i.e., that $P(\epsilon, p_T)$ is negligible unless $\epsilon \left /\right. p_T \ll 1$, and that $n$ is sufficiently large ($n \sim 5$). The result is then written to first order in $\langle \epsilon \rangle (p_T) \left /\right. p_T$, with $\langle \epsilon \rangle (p_T) = \int_0^{+\infty} \, d\epsilon \, P(\epsilon, p_T) \, \epsilon$. 

The upper cumulative jet cross sections in Eq.~\eqref{eq:quantile_matching} read
\begin{align}
    \Sigma^{pp}(p_T^{v},+\infty) & = 
    \int_{p_{T}^{v}}^{+\infty} dp_T \frac{d\sigma^{pp}}{dp_T} =
    \frac{A}{n-1}(p_T^{v})^{1-n}\, , \label{cumulative_xs_pp} \\
    \Sigma^{AA}(p_T^{q},+\infty) & = 
    \int_{p_{T}^{q}}^{+\infty} dp_T \frac{d\sigma^{AA}}{dp_T}
    \approx \frac{A}{n-1}\left(p_T^{q} + \langle \epsilon \rangle (p_T^q)\right)^{1-n}\, , \label{cumulative_xs_AA}
\end{align}
where we further assumed that $\langle \epsilon \rangle(p_T)$ varies slowly with $p_T$ compared to $p_T^{-n}$ and that $n$ is constant. Inserting Eqs.~\eqref{cumulative_xs_pp} and~\eqref{cumulative_xs_AA}  into Eq.~\eqref{eq:quantile_matching}, we have
\begin{align}\label{eq:qaa_estimate}
    1-\frac{p_T^q(p_T^v)}{p_T^v} = 1-Q_{AA}(p_T^v) \approx  \frac{\langle \epsilon \rangle  (p_T^v)}{p_T^v}\, ,
\end{align}
where we took $\langle \epsilon \rangle (p_T^q) \approx \langle \epsilon \rangle (p_T^v)$, i.e., the zeroth order term in the expansion of $\langle \epsilon \rangle$ in powers of $(p_T^v-p_T^q)$. This confirms that $1-Q_{AA}$ is a proxy for the \textit{average} fractional energy loss as a function of $p_T$. This proxy is independent of the spectrum slope when assuming a $p_T^{-n}$ shape with constant $n$ and assumes a small relative energy loss with a mild $p_T$ dependence.\\

One can also formulate a local version of the integral equation in Eq.~\eqref{eq:quantile_matching} by differentiating it with respect to $p_T^v$\\
\begin{equation}\label{eq:diff_qaa}
    \frac{dp_T^q}{dp_T^v}\frac{d\sigma^{AA}}{dp_T^q}(p_T^q(p_T^v)) = \frac{d\sigma^{pp}}{dp_T^v}(p_T^v) \, ,
\end{equation}
which results in a first order differential equation for $p_T^q(p_T^v)$. A calculation of $p_T^q(p_T^v)$ based on Eq.~\eqref{eq:diff_qaa} is potentially more sensitive to statistical uncertainties in measured spectra. 
Additionally, one should keep in mind that, although Eq.~\eqref{eq:diff_qaa} does not need an integration up to arbitrarily large $p_T$ in contrast with Eq.~\eqref{eq:quantile_matching} (see Section \ref{sec:cutoff}), it is, in principle, still sensitive to a choice of $p_T^q$ evaluated at some momentum. This is because Eq.~\eqref{eq:diff_qaa} is a first order differential equation in $p_T^q$ for which an initial condition needs to be specified.

Interestingly, the left-hand side of Eq.~\eqref{eq:diff_qaa} was first introduced in \cite{PHENIX:2004vcz} in the context of obtaining a simple energy loss estimate by looking at the suppression of the $\pi^0$ spectrum\footnote{Note that, in Eq.~(29) of \cite{PHENIX:2004vcz}, the AA spectrum differential in vacuum momentum $p_T'$ and evaluated at a medium momentum $p_T$ ($dn/p_T'$ in that work) is taken as equal to the pp spectrum differential in $p_T'$ evaluated at $p_T'$ (i.e. $\propto 1 \, / \, (p_T')^{n-1}$ in that work). This is exactly the statement of the quantile procedure, as one can directly verify by looking at its differential version in Eq.~\eqref{eq:diff_qaa}.}. It was used to quantify the number of particles in a bin of $p_T$ after suppression. The jacobian factor $dp_T^q \left/ \right. dp_T^v$ accounts for the change in the relative particle density per $p_T$ bin as a consequence of induced energy loss. Calculating $p_T^q$, whether via Eq.~\eqref{eq:diff_qaa} or Eq.~\eqref{eq:quantile_matching}, is a natural extension of this logic, in that it imposes that the number of in-medium jets in a given $p_T^q$ bin scaled by the jacobian should be the same as the number of vacuum jets in a given $p_T^v$ bin when we assume jets to be ideally ordered as described above. Note that a relation similar to Eq.~\eqref{eq:diff_qaa} was used in \cite{ATLAS:2023iad} as an experimental proxy for jet energy loss. The two coincide in the limit where  $dp_T^q \left/ \right. dp_T^v \rightarrow 1$, i.e., slowly varying energy loss with $p_T$, in which case the two equations determine similar quantities (modulus the assignment $p_T^q(p_T^v) = p_T^v - \Delta p_T(p_T^v) = p_T^v(1 - S_{loss}(p_T^v)$). We explore this further in Section \ref{subsec:sloss_qaa}.
The equation used in \cite{ATLAS:2023iad} is a differential equation and thus, like Eq.~\eqref{eq:diff_qaa}, also needs an initial condition to be specified\footnote{We were unable to identify the initial condition chosen in \cite{ATLAS:2023iad}.}.


\section{Generation and analysis details}
\label{sec:generation}

The samples used in this work were generated with  \jewel\ 2.3 \cite{Zapp:2012ak, Zapp:2013vla}, and with the Hybrid Model\footnote{We thank the authors of~\cite{Bossi:2024qho} for kindly providing us with these samples.} \cite{Casalderrey-Solana:2014bpa,Casalderrey-Solana:2015vaa,Casalderrey-Solana:2016jvj,Hulcher:2017cpt,Casalderrey-Solana:2018wrw} for model comparison of part of the results.

\jewel\ is a Monte Carlo event generator that implements jet evolution both in vacuum and in a medium, describing the QCD evolution of hard partons and their scatterings with the medium in a perturbative framework. 
\jewel\ interleaves the final state parton shower with scatterings with a QCD medium with leading-log correct relative contributions for elastic and inelastic scattering, such that vacuum-like and medium-induced emissions are treated in the same perturbative language. The interplay of both types of radiation is governed by formation times -- all emissions are due to happen, but the one with the longest formation time is discarded. With this construction and using virtuality as an ordering variable, radiative energy loss is included while accounting for the usual vacuum parton shower. Additionally, \jewel\ implements destructive interference between emissions induced by successive medium scatterings via a probabilistic formulation of the QCD Landau-Pomeranchuk-Migdal (LPM) effect.

\jewel\ simulates the medium as the boost-invariant longitudinal expansion of an ideal quark-gluon gas. 
\jewel\ has the option of storing the four-momenta of the partonic constituents of the medium that were scattered (recoils) and also their initial stage (thermal) counterparts in the event record. This information is used to perform a background subtraction bettering the quantitative understanding of medium response and allowing for the reconstruction of a heavy-ion jet that takes into account medium response while discarding uncorrelated background. In this work, we use the event-wise algorithmic procedure introduced in \cite{Milhano:2022kzx} to subtract the background prior to jet reconstruction. The cutoff distance for subtraction is set to $\Delta R_{ij} = 0.5$. All remaining \jewel\  parameters concerning recoiling partons are set to their default values.

In the Hybrid Model \cite{Casalderrey-Solana:2014bpa,Casalderrey-Solana:2015vaa,Casalderrey-Solana:2016jvj,Hulcher:2017cpt,Casalderrey-Solana:2018wrw}, jet evolution inside a droplet of QGP is described by separating the strongly coupled dynamics of QGP and of the jet-QGP interactions, both governed by the medium temperature scale $T$, from the weakly coupled dynamics governing the production and showering of high energy partons. Since the initial virtuality of the parton that showers and forms a jet is much larger than any scale associated with the medium (e.g. its temperature), the model assumes that its evolution proceeds as in vacuum, i.e., the parton shower branching is unmodified by the presence of the plasma. Specifically, the evolution is given by the parton shower implemented in PYTHIA 8.244~\cite{Sjostrand:2014zea} and includes initial state radiation. As a consequence, the modification of jets is only due to the interaction of each parton in the jet with the strongly coupled medium. The rate of parton energy loss is modeled by the energy loss of light quark jets in the strongly coupled plasma of $N = 4$ supersymmetric Yang-Mills theory (SYM) which has been computed holographically~\cite{Chesler:2014jva,Chesler:2015nqz}. This rate is applied to each parton of the already developed shower, using a formation time estimate for the path length travelled by each of them. Moreover, it is assumed that the all the differences between the strongly coupled limit of $N = 4$ SYM theory and QCD can be enconded into a single parameter which is fitted to jet and hadron $R_{AA}$ data \cite{Casalderrey-Solana:2018wrw}.

Each event is embedded within a droplet of QGP which expands and cools as described by relativistic viscous hydrodynamics, providing an evolving space-time background which is averaged over collisions within a given centrality class. The holographic energy loss rate is computed in the local rest frame of this fluid and at its local temperature. The jet propagates through the expanding cooling medium until it reaches a region where the temperature has dropped below a critical temperature $T_c$ at which point no further energy loss occurs. The value for this temperature is varied within a given range $145 < T_c < 170$ MeV, gauging theory uncertainties of the model. 

The energy and momentum deposited by the energetic partons of the shower in the medium excites a wake, which then evolves hydrodynamically. When the QGP and the jet-induced wake reach the freeze-out temperature, they are hadronized via the the Cooper-Frye prescription~\cite{PhysRevD.10.186}. This prescription ensures that by boosting the fluid in its wake in its direction, the jet enhances the production of soft particles in its direction and depletes it in the opposite direction, relative to what would be obtained for the case without any jet wake. In the Hybrid Model, where the resulting Cooper-Frye distribution 
is negative it is implemented by adding hadrons with this distribution but with a negative energy. These "negative particles" are subtracted by using the same procedure employed for the thermal momenta in \jewel.
~\\

\begin{figure}
     \centering
     \begin{subfigure}[h]{0.49\textwidth}
         \centering
         \includegraphics[width=\textwidth]{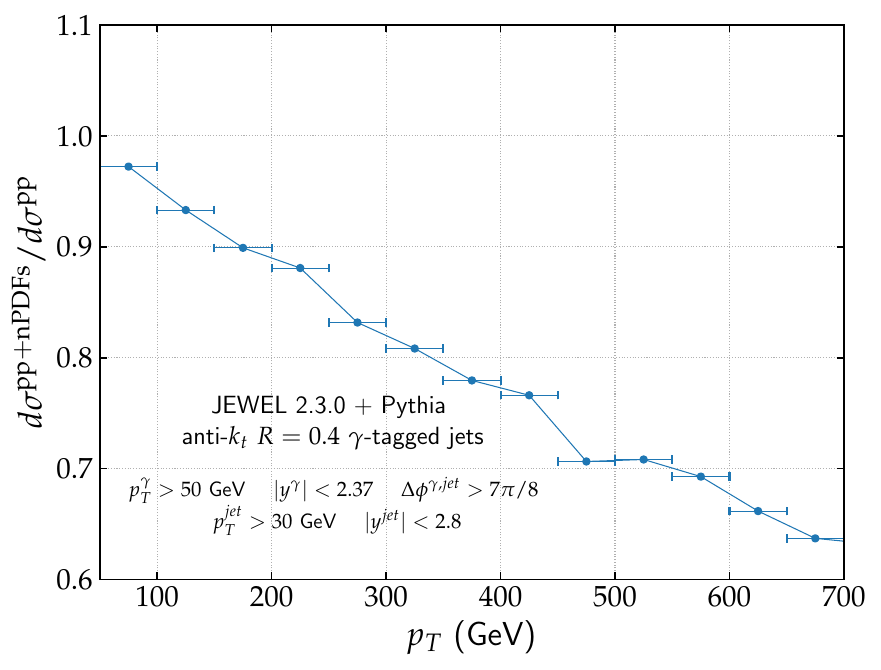}
         \label{dfd}
     \end{subfigure}
     \begin{subfigure}[h]{0.49\textwidth}
         \centering
         \includegraphics[width=\textwidth]{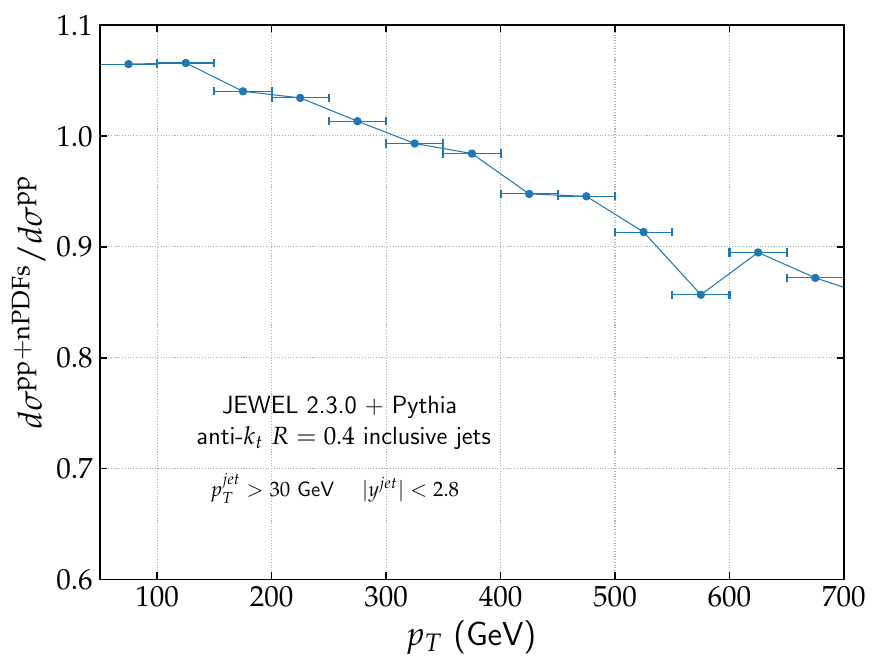}
         \label{fig:three sin x}
     \end{subfigure}
     \begin{subfigure}[h]{0.49\textwidth}
         \centering
         \includegraphics[width=\textwidth]{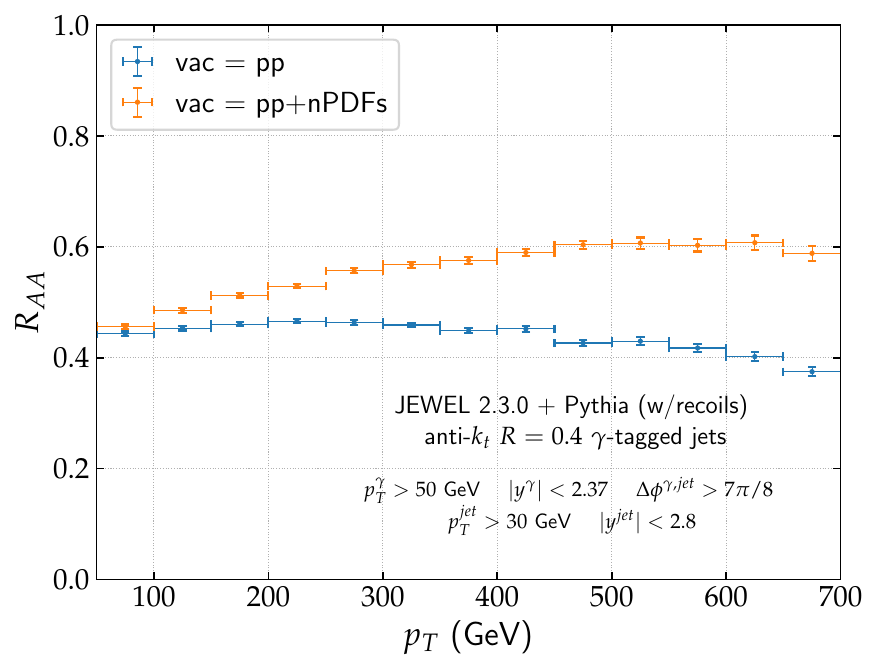}
         \label{dfd}
     \end{subfigure}
     \begin{subfigure}[h]{0.49\textwidth}
         \centering
         \includegraphics[width=\textwidth]{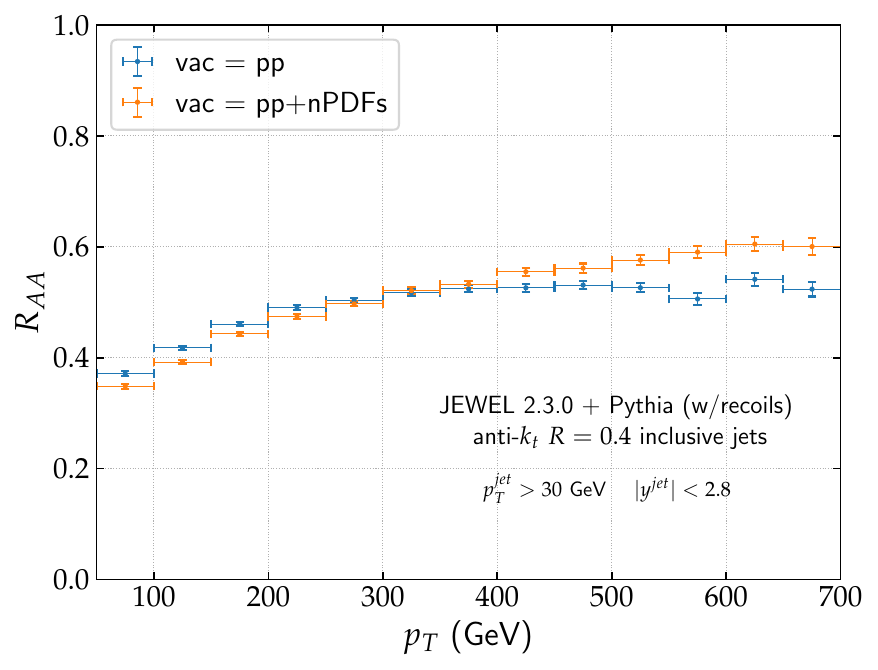}
         \label{fig:three sin x}
     \end{subfigure}
     \caption{\textbf{Top:} Sensitivity of the vacuum $\gamma$-jets (\textbf{left}) and inclusive jets (\textbf{right}) spectra to nuclear effects. \textbf{Bottom:} Sensitivity of $R_{AA}$ to nuclear effects. The spectra labeled with pp+nPDFs result from (average)nucleon-(average)nucleon collisions generated with isospin averaged nuclear PDFs.}
     \label{fig:nucleareffects}
\end{figure}

\jewel\ samples were generated for $\sqrt{s} = 5.02$ TeV, for both pp and PbPb ($[0-10]\%$ centrality) collisions. These samples consist of $10^6$ dijet events  and $10^6$ $\gamma+$jet events. 
For both dijet and $\gamma+$ jet events, the 
EPPS16NLO nuclear PDFs \cite{nPDFs} (nPDFs) are used for PbPb collisions. For dijet events, for the most part of the work, the CT14NLO proton PDFs \cite{PDFs} are used for pp collisions. All sets are provided by LHAPDF \cite{lhapdf}. The particles produced in the hard scattering have a minimum transverse momentum cutoff of $20$ GeV and the medium is generated with $\tau_i = 0.6$ fm, $T_i = 0.55$ GeV and $T_c = $ 0.17 GeV. Initial state radiation and hadronization are handled by the PYTHIA 6~\cite{Sjostrand:2006za} framework that underlies \jewel. 

Importantly, pp samples for $\gamma+$ jet events are generated with the EPPS16NLO nPDFs, which are isospin averaged, as initial conditions. Hence, instead of the usual pp collisions, vacuum samples are obtained from the unphysical scenario of (average)nucleon-(average)nucleon collisions with no QGP. This is an attempt at isolating the role of quenching effects in the differences between vacuum and medium samples by minimizing the role of nuclear effects (isospin~\cite{Li:2019dre} and nPDFs), which is significant for $\gamma+$jet events \cite{ATLAS:2023iad}. 
Note that, for PbPb events, both \jewel\, and the Hybrid Model only generate a single (average)nucleon-(average)nucleon collision per event. The cross-section estimate per event already takes this into account, since it is weighted by the isospin averaged nuclear PDFs. To illustrate the role of nuclear effects on the inclusive jet  and $\gamma+$jet spectra we show, in the top panels of Fig.~\ref{fig:nucleareffects}, the ratio between vacuum jet spectra obtained from events generated with isospin averaged nPDFs and events generated with proton PDFs. It is clear that the nuclear effects are non-negligible both in absolute magnitude of the cross-section and in the shape of the spectrum. This is especially pronounced for $\gamma+$jets, where the isospin effect in particular is expected to be greater since, at leading-order, the hard matrix elements always include an initial state quark or anti-quark. On the bottom panels of Fig.~\ref{fig:nucleareffects} we show the impact of nuclear effects on an observable like the jet $R_{AA}$. At large $p_T$ and for $\gamma+$jets, this translates into a factor $\sim1.5$ difference. Since nuclear effects are not as significant for dijet events, we will, for the most part, use dijet vacuum samples obtained from simple pp collisions. This is true except for when drawing comparisons with $\gamma+$jet results in Section~\ref{sec:color_carge}, for consistency.

The Hybrid Model samples were generated for $\sqrt{s} = 5.02$ TeV, for both pp and PbPb ($[0-5]\%$ centrality) collisions. These samples consist of $\sim 10^6$ dijet events. The EPS09LO nuclear PDFs~\cite{Eskola:2009uj} are used for PbPb collisions and the NNPDF2.3 PDFs for pp collisions. Importantly, the particles produced in the hard scattering have a minimum transverse momentum cutoff of $100$ GeV. For more details on the generated samples see~\cite{Bossi:2024qho}.
~\\

Let us specify the details of this work's analysis. In $\gamma+$jet events, the hard scattered photon is identified solely based on its $p_T$, i.e., the highest $p_T$ photon from each event which passes $p_T \geq 50$ GeV and $|y| \leq 2.37$ is chosen. Note that \jewel\ does not implement QED radiation off either initial- or final-state quarks and, therefore, most photons not coming from the hard scattering come from the $\pi^0$ decay. The photon tagging is done prior to subtracting the background in PbPb events, keeping the photon out of the process. After the subtraction procedure, for both types of events, the final state particles are filtered according to $p_T \geq 0.5$ GeV (unless otherwise stated as in Section~\ref{sec:ptmin}) and $|y| \leq 3.3$ and then given as input to the anti-$k_t$ algorithm \cite{Cacciari:2008gp} for jet reconstruction with varying jet radius, as implemented in \texttt{FASTJET} \cite{fastjet}. All jets have a minimum cut of $p_T^{jet} \geq 30$ GeV ($\geq 130$ GeV) in \jewel\ (Hybrid Model) and are required to have $|y^{jet}| \leq 2.8$. The $\gamma$-tagged jets are required to be azimuthally separated by $\Delta\phi \geq 7\pi/8$ from the hard photon. For results using dijet events, the jet sample is inclusive.

\section{Validation of the quantile procedure as a proxy for average energy loss}\label{sec:qaa_validation}

First, we validate the quantile procedure as an adequate proxy for jet energy loss away from the zero dispersion (and ideally ordered) case described in Section \ref{sec:qaa_intro}. We do this by checking that the quantile procedure gives the same information as the $p_T$ of the photon does in $\gamma$+jet events. Events with such a topology are particularly useful when validating any type of energy loss claim given that the transverse momentum of the photon is a proxy for the transverse momentum of the parton that initiated the jet found back-to-back with it. Given that nuclear effects (isospin and nuclear PDF) significantly alter the shape of the PbPb spectrum for $\gamma+$jet events, vacuum samples are generated with nuclear PDFs, which are isospin averaged, as initial conditions (see Section \ref{sec:generation}).

\begin{figure}
     \centering
     \begin{subfigure}[h]{0.49\textwidth}
         \centering
         \includegraphics[width=\textwidth]{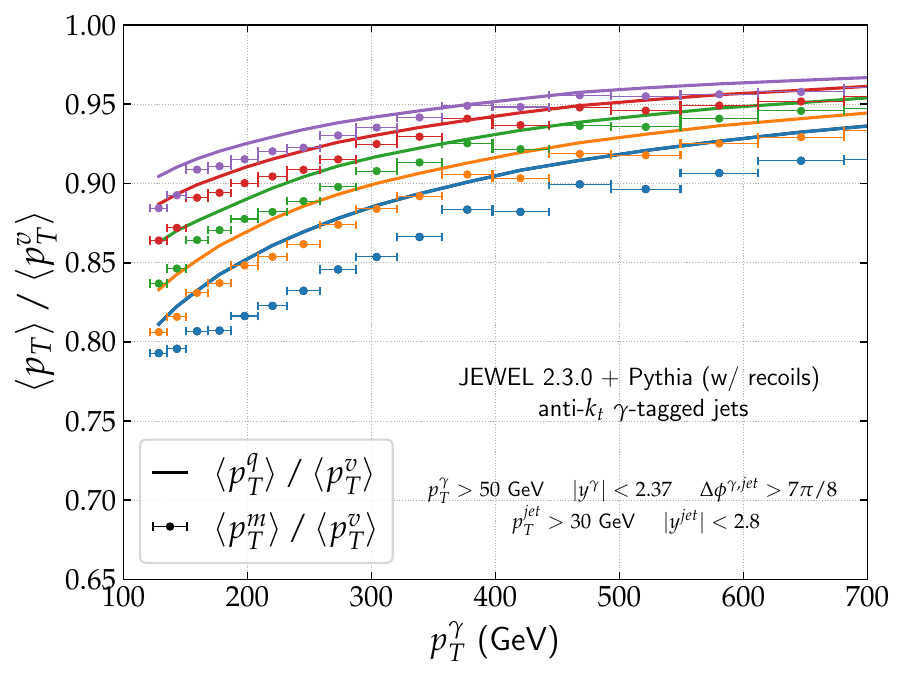}
         \label{dfd}
     \end{subfigure}
     \begin{subfigure}[h]{0.49\textwidth}
         \centering
         \includegraphics[width=\textwidth]{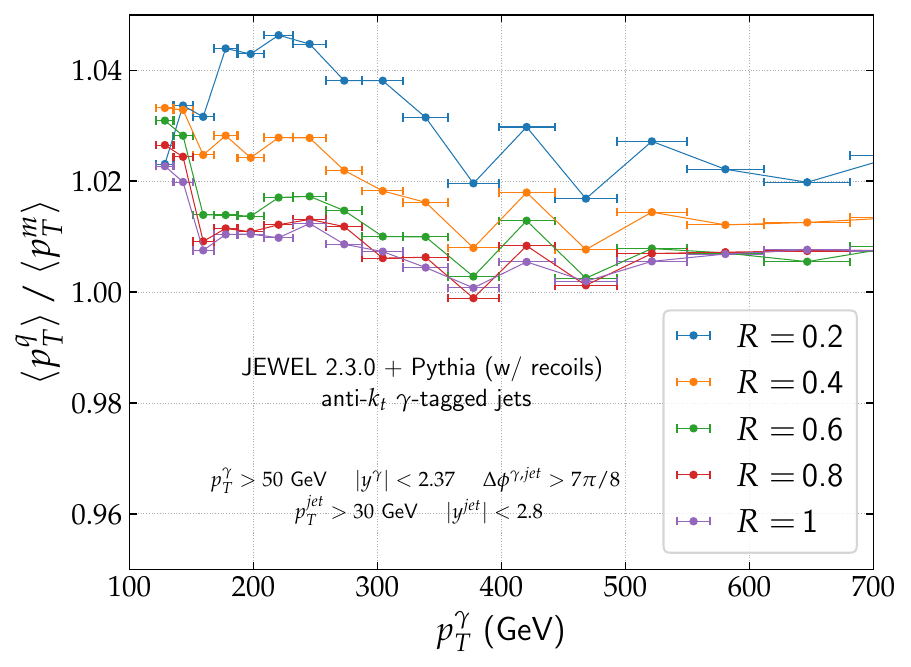}
         \label{fig:three sin x}
     \end{subfigure}
     \caption{\textbf{Left}: Dependence of average jet energy loss proxies on jet radius $R$ -- ratio of mean values of reconstructed $p_T$ (dots) and ratio of average quantile $p_T$ to mean vacuum $p_T$ (solid). \textbf{Right}: Ratio between mean $p_T^q$ and mean $p_T^m$. Vacuum samples include nuclear effects (isospin averaged nuclear PDFs).}
     \label{fig:qaa_validation_r}
\end{figure}

In \cite{Brewer:2018dfs}, the authors confirmed the validity of the quantile procedure for $R=0.4$ jets produced back-to-back with a $Z$ boson in the absence of medium response contributions to the jet. Here, we extend the validation to the more realistic case where medium response is accounted for (as described in Section \ref{sec:generation}) and for multiple jet radii. 
The reasoning for this validation goes as follows. For a given photon $p_T^{\gamma}$ produced in the hard scattering, a distribution of jet $p_T$, $p_T^v$ for vacuum samples and $p_T^m$ for medium samples, is produced. The ratio between the mean value of these distributions $\langle p_T^m \rangle \left / \right.\langle p_T^v \rangle$ (solid curves on the left plot of Fig.~\ref{fig:qaa_validation_r}) gives us a proxy for average jet energy loss, given that we are comparing jets originated by hard partons with the same $p_T$ but that developed differently (vacuum and medium). Note that this is not the same as the average of the ratio of these two quantities, which is inaccessible in both \jewel\ and in experimental data. 
The mean $p_T$ values are given by 
\begin{equation}\label{eq:av_ptv}
    \left.\langle p_T^{v/m}\rangle\right|_{p_T^{\gamma}} = \int dp_T \left.\frac{dN^{v/m}}{dp_T}\right|_{p_T^{\gamma}} p_T \,,
\end{equation}
where $\left.\frac{dN^X}{dp_T}\right|_{p_T^{\gamma}}$ is the normalized transverse momentum distribution of jets for a given bin of photon transverse momentum $p_T^{\gamma}$. 
On the other hand, the quantile procedure determines $p_T^q = p_T^q(p_T^v)$ for a given $p_T^v$, which should relate pp and AA jets that, on average, were originated by hard partons with the same $p_T$.
One can then take the distribution of vacuum jets for a given $p_T^{\gamma}$ and calculate an hypothetical medium $p_T$ distribution from it by applying $p_T^q(p_T^v)$. The average of this distribution is then given by
\begin{equation}\label{eq:av_ptq}
    \left.\langle p_T^{q}\rangle\right|_{p_T^{\gamma}} = \int dp_T \left.\frac{dN^{v}}{dp_T}\right|_{p_T^{\gamma}} p_T^{q}(p_T) \,.
\end{equation}
The ratio $\langle p_T^q \rangle \left / \right. \langle p_T^v \rangle$ (solid curves in Fig.~\ref{fig:qaa_validation_r}) is a proxy for the average jet energy loss for a given photon $p_T^{\gamma}$, which should be similar to $\langle p_T^m \rangle \left / \right.\langle p_T^v \rangle$ (dots in Fig.~\ref{fig:qaa_validation_r}).
Note, again, that this is not the same as the mean ratio of $p_T$ values. In this case, that would correspond to an average value of $Q_{AA}$, which, although very similar to the quantile proxy we show here, is not the right quantity for this validation. With this being said, on the left-hand plot of Fig.~\ref{fig:qaa_validation_r}, we see that the agreement between the two average jet energy loss proxies is reasonable, although the quantile procedure systematically results in a smaller average energy loss than that estimated using the photon $p_T$. 

On the right-hand plot of Fig.~\ref{fig:qaa_validation_r}, we directly show the comparison between $\langle p_T^{q}\rangle$ and $\langle p_T^{m}\rangle$. A quantile procedure which perfectly quantified the correspondence between mean medium and vacuum jet $p_T$ would result in a ratio of exactly $1$. This ratio approaches $1$ for larger the jet radii and for larger jet $p_T$, saturating at about a $1\%$ difference. 
The largest deviation, which does not exceed $\sim 5\%$, is for $R=0.2$ jets.
We that note that medium response is a large source of fluctuations pushing us away from the ideally ordered scenario. 
Looking at the left-hand plot of Fig.~\ref{fig:qaa_validation_r}, one can also see how jet energy loss varies with jet radius, for radii from $0.2$ to $1$. 
By using information from the photon’s $p_T$, we could already see (solid curves) that the ratio $\langle p_T^m \rangle \left / \right. \langle p_T^v \rangle$ increases as one increases $R$. Although this is a model dependent statement (see Section~\ref{subsec:jet_radius} for model comparison), this means that larger jets lose a smaller fraction of their energy, on average. One can interpret this as due to medium response and to the recapture of medium-induced radiation compensating for the fact that increasing the jet radius increases the number of jet components that can lose energy. Both proxies have the same evolution with jet radius.

\section{Jet energy loss dependence on jet radius, minimum particle $p_T$ and colour charge}\label{sec:results}
In the following subsections, we use $Q_{AA}$ as a proxy for average fractional jet energy loss and investigate its evolution with jet radius, its sensitivity to medium response and minimum particle $p_T$ and, lastly, use it to evaluate the colour charge dependence of energy loss. Only on this last study are $\gamma+$jet samples used. For the remaining two, the results are based on inclusive jet samples.

\subsection{Jet radius}\label{subsec:jet_radius}

\begin{figure}[h]
     \centering
     \begin{subfigure}[h]{0.49\textwidth}
         \centering
         \includegraphics[width=\textwidth]{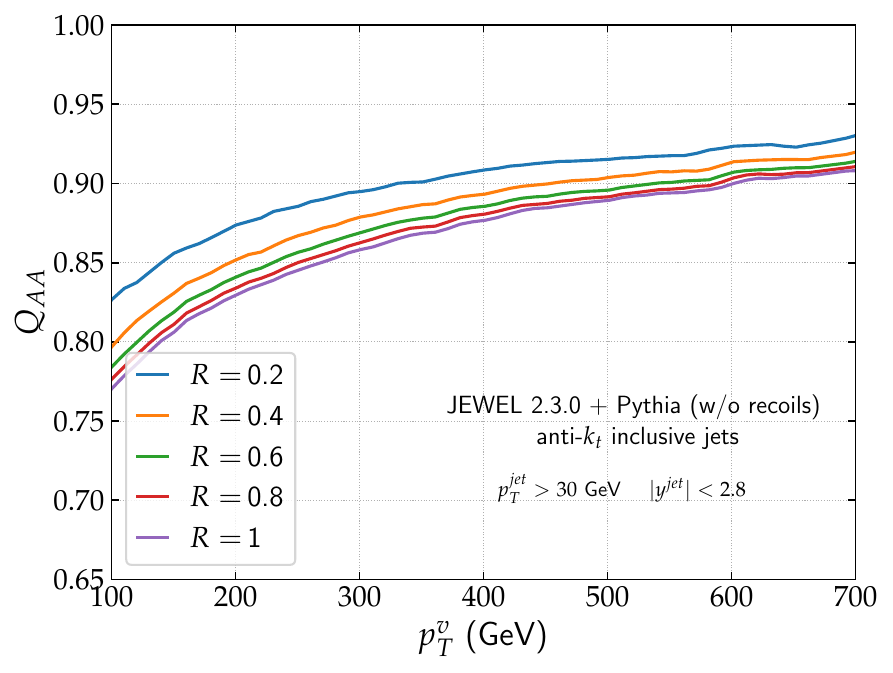}
         \label{dfd}
     \end{subfigure}
     \begin{subfigure}[h]{0.49\textwidth}
         \centering
         \includegraphics[width=\textwidth]{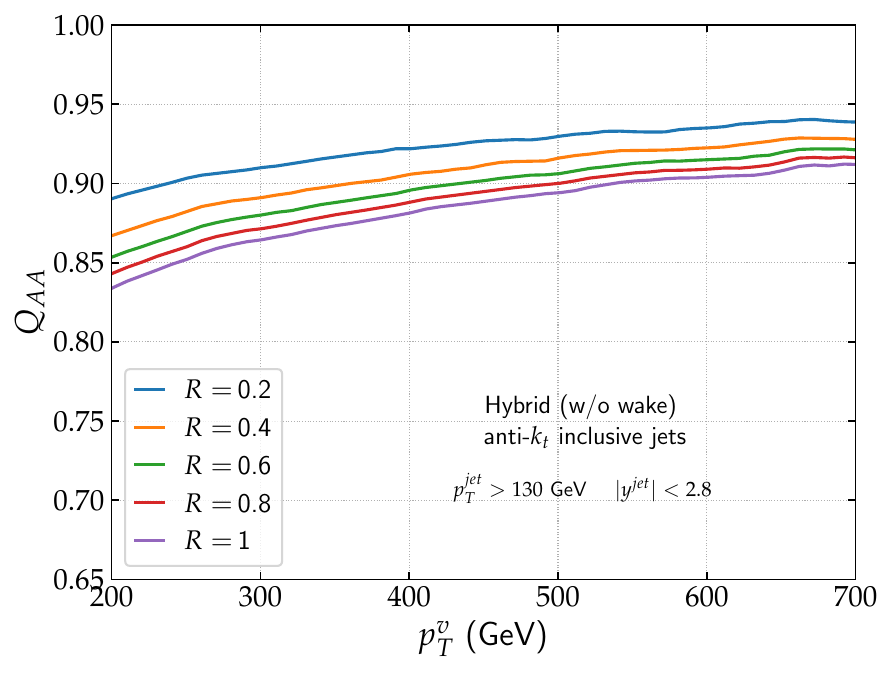}
         \label{fig:three sin x}
     \end{subfigure}
     \begin{subfigure}[h]{0.49\textwidth}
         \centering
         \includegraphics[width=\textwidth]{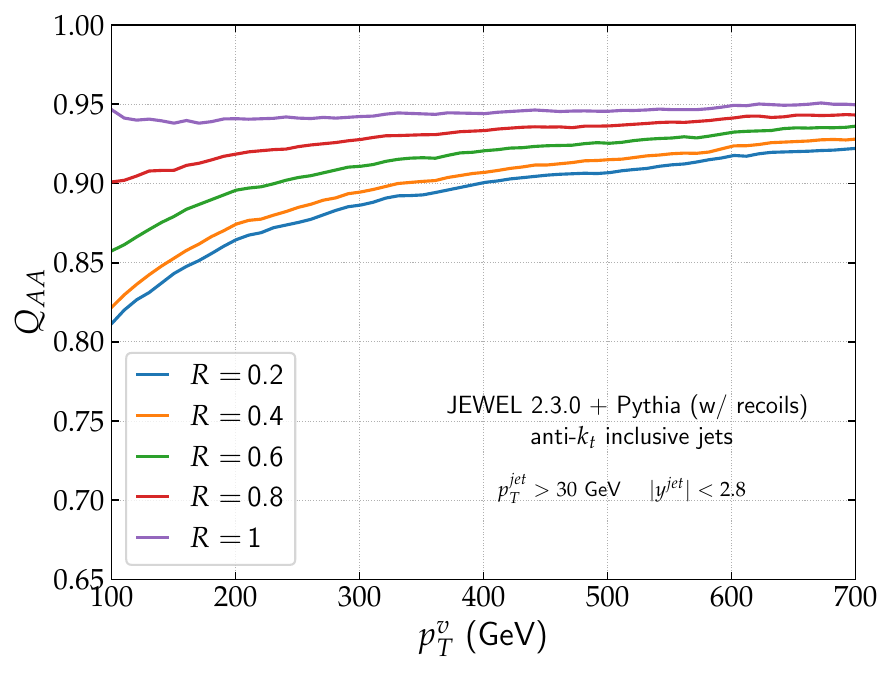}
         \label{dfd}
     \end{subfigure}
     \begin{subfigure}[h]{0.49\textwidth}
         \centering
         \includegraphics[width=\textwidth]{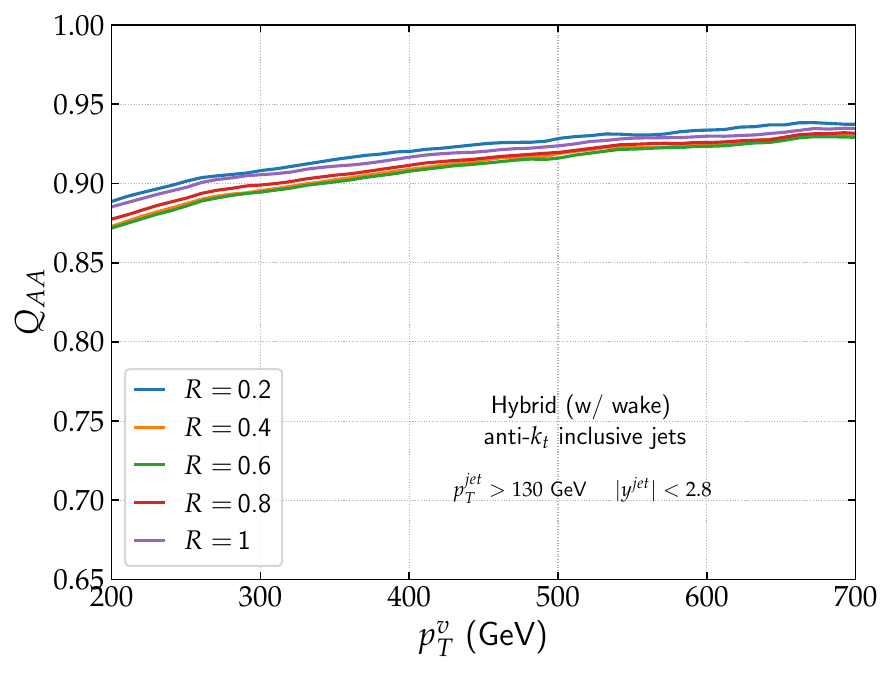}
         \label{fig:three sin x}
     \end{subfigure}
     \caption{Dependence of $Q_{AA}$ with jet radius for \jewel\ 2.3 (\textbf{left}) and for the Hybrid Model (\textbf{right}), with (\textbf{bottom}) and without (\textbf{top}) medium response. Vacuum samples come from pp collisions.}
      \label{fig:qaa_model}
\end{figure}
In Fig.~\ref{fig:qaa_model} we show the evolution of $Q_{AA}$ with jet radius for inclusive jet samples with (bottom panels) and without (top panels) medium response. For the purpose of evaluating model dependence, we compare two radically different implementations of jet physics in the presence of QGP - \jewel\ 2.3 \cite{Zapp:2012ak, Zapp:2013vla} (left) and the Hybrid Model \cite{Casalderrey-Solana:2014bpa,Casalderrey-Solana:2015vaa,Casalderrey-Solana:2016jvj,Hulcher:2017cpt,Casalderrey-Solana:2018wrw} (right). Note that the curves for the Hybrid Model (right) start at $200$ GeV. 
Despite the difference in parton-QGP interaction in these models, the trend of $Q_{AA}$ with jet radius and $p_T$ is the same when considering events without medium response (top panels in Fig~\ref{fig:qaa_model}). Even in absolute value the two models predict similar results. The ordering of $Q_{AA}$ with jet radius is such that larger jets lose a larger fraction of their energy, on average. This implies, in particular for \jewel, that the increase in jet components that can lose energy dominates over the recapture of medium-induced radiation. 
When including medium response (bottom panels in Fig.~\ref{fig:qaa_model}), however, the models predict different orderings with jet radius and only agree approximately in absolute value for large enough $p_T$ and moderate radii $R\sim 0.4,0.6$. For \jewel\ the behaviour with jet radius is exactly inverted with respect to the case where medium response is absent, which is naturally attributed to the reconstruction of part of the energy lost to the medium inside the jet. This had already been observed for jet suppression ($R_{AA}$) in \cite{KunnawalkamElayavalli:2017hxo}. Nevertheless, the results we show for $Q_{AA}$ allow for this statement to be made quantitatively with regards to average jet energy loss. As for the Hybrid model, the rate of energy recovery by reconstructing part of the wake inside the jet is not very pronounced. Thus, for small to moderate jet radii, the energy recovery from the wake does not outgain the increase in lost energy due to the increase in number of jet components. It is only for large enough radii $R \gtrsim 0.6$ that this happens, causing the ordering with jet radius to invert like in \jewel.
Furthermore, with regards to the sensitivity of $Q_{AA}$ to the jet radius, in the absence of medium response, \jewel\ predicts a similar radius dependence to the Hybrid Model's for jet radii $R \lesssim 0.6$. For larger jet radii, the dependence is milder for \jewel. When medium response is included, the radius dependence of \jewel\ is pronouncedly larger than that of the Hybrid Model. This difference in behaviours can be explained by different rates of recapture of energy. Without medium response, jets in Hybrid Model have no way of recapturing lost energy, such that increasing the jet radius only increases the number of components of the vacuum parton shower that experience energy loss. On the other hand, for \jewel, radius dependence starts saturating at intermediate jet radii ($R \sim 0.6$) because of the recapture of medium-induced emissions. When including medium response, \jewel\ jets with increasing radius, recapture energy at a significantly faster rate than in the Hybrid Model\footnote{Note that \jewel's rate for energy recovery may be exaggerated due to the value of the subtraction parameter $\Delta R_{ij}$ being set to $0.5$. For jet radii sufficiently larger than this value, one may be undersubtracting. In fact, we have checked that this rate is slowed down when increasing $\Delta R_{ij}$ to $1$. Nevertheless, all statements regarding \jewel\ and Hybrid comparisons in this and in the following section remain unaltered.}.
In fact, in \cite{Milhano:2022kzx} it was shown that \jewel's medium response is overpopulated in semi-hard portion of the particle spectrum, possibly a consequence of the recoiling medium particles not suffering any type of shower evolution or re-scattering. In contrast, in the Hybrid Model, medium response is composed of mostly very soft particles and spreading to large distances \cite{Casalderrey-Solana:2016jvj}. This means that, not only is the portion of lost energy that the jet can recover from the wake limited, but also that one needs to increase the jet radius significantly to be able to recapture a substantial amount of that energy.

\subsection{Minimum particle $p_T$}\label{sec:ptmin}
\begin{figure}
     \centering
     \begin{subfigure}[h]{0.49\textwidth}
         \centering
         \includegraphics[width=\textwidth]{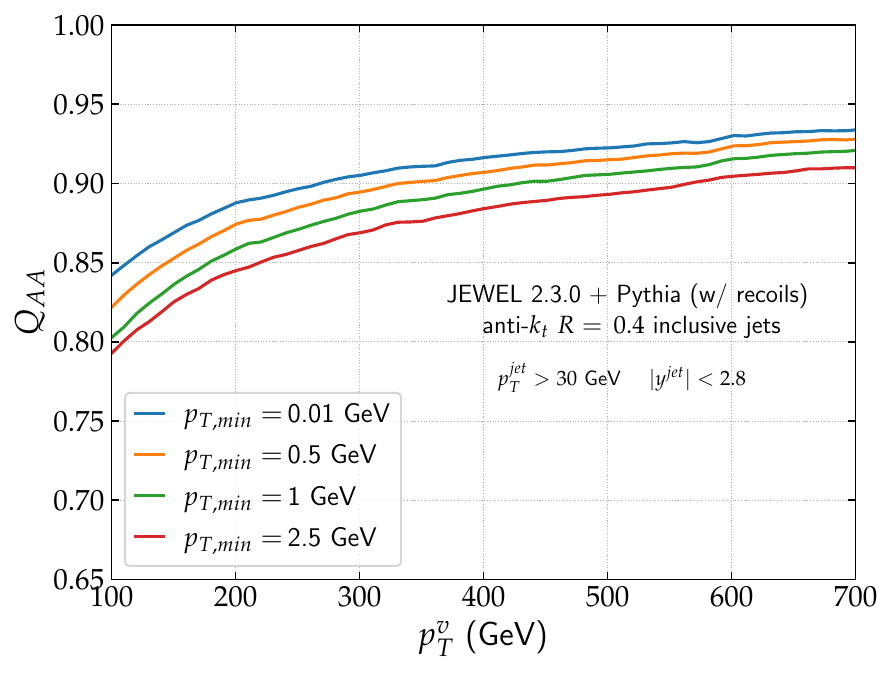}
         \label{dfd}
     \end{subfigure}
     \begin{subfigure}[h]{0.49\textwidth}
         \centering
         \includegraphics[width=\textwidth]{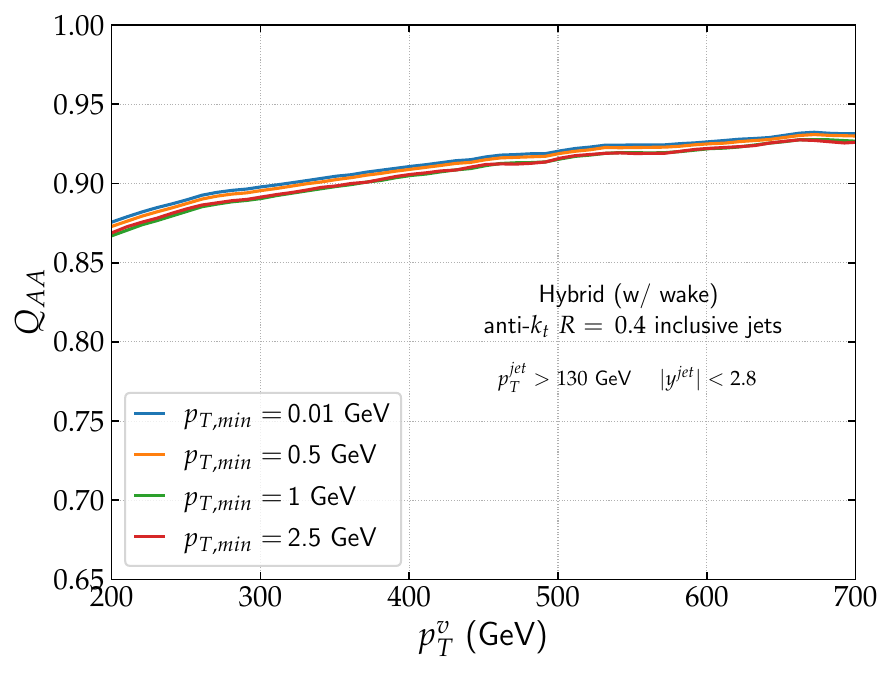}
         \label{dfd}
     \end{subfigure}
     
     \begin{subfigure}[h]{0.49\textwidth}
         \centering
         \includegraphics[width=\textwidth]{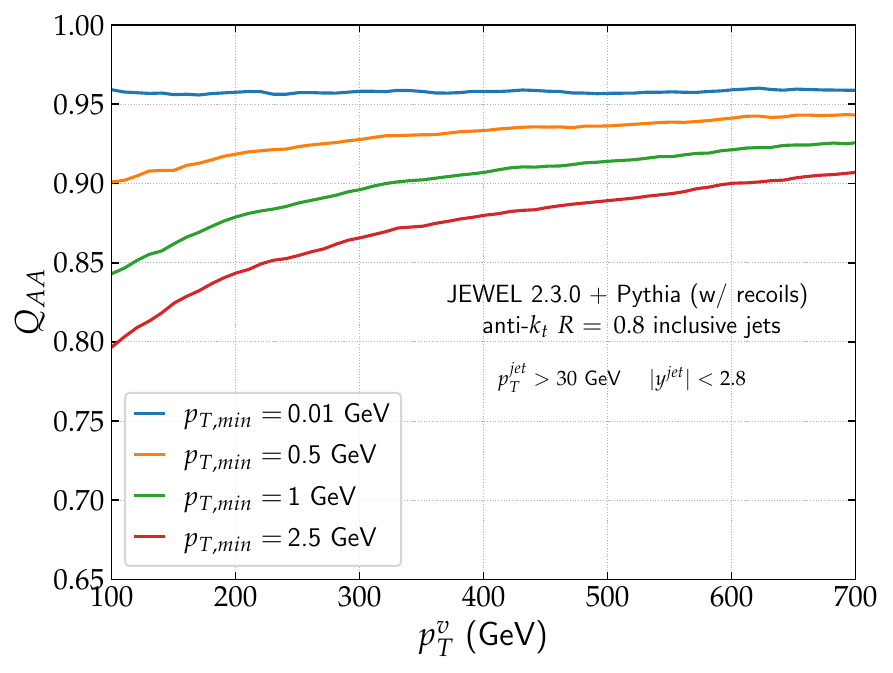}
         \label{fig:three sin x}
     \end{subfigure}
     \begin{subfigure}[h]{0.49\textwidth}
         \centering
         \includegraphics[width=\textwidth]{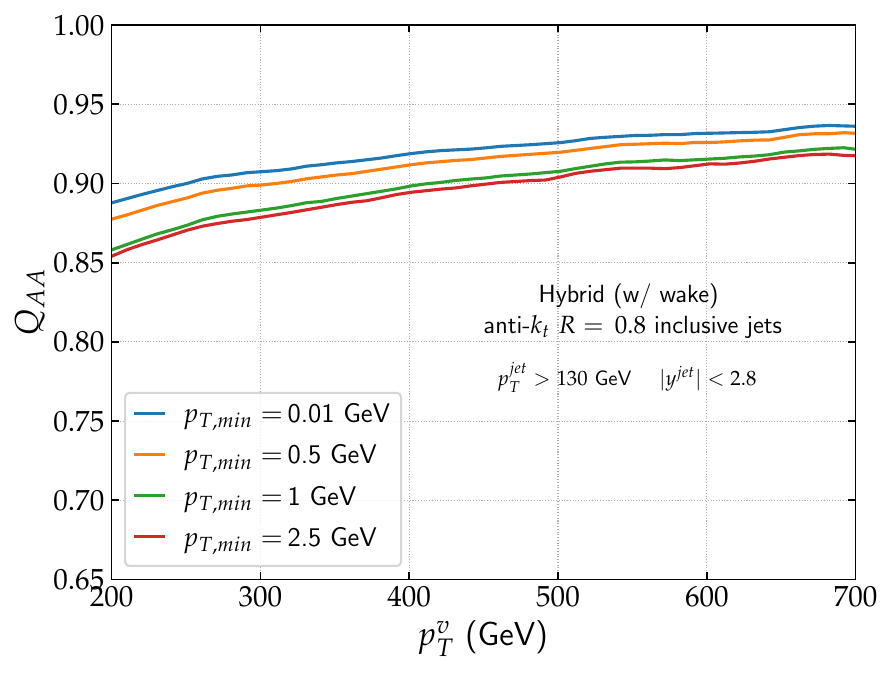}
         \label{fig:three sin x}
     \end{subfigure}

      \begin{subfigure}[h]{0.49\textwidth}
         \centering
         \includegraphics[width=\textwidth]{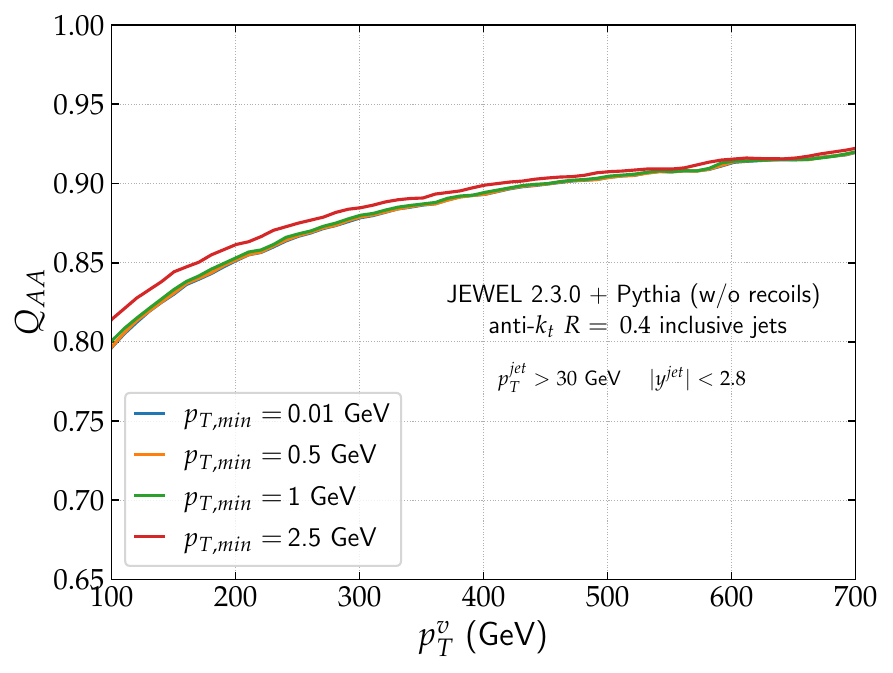}
         \label{fig:three sin x}
     \end{subfigure}
     \begin{subfigure}[h]{0.49\textwidth}
         \centering
         \includegraphics[width=\textwidth]{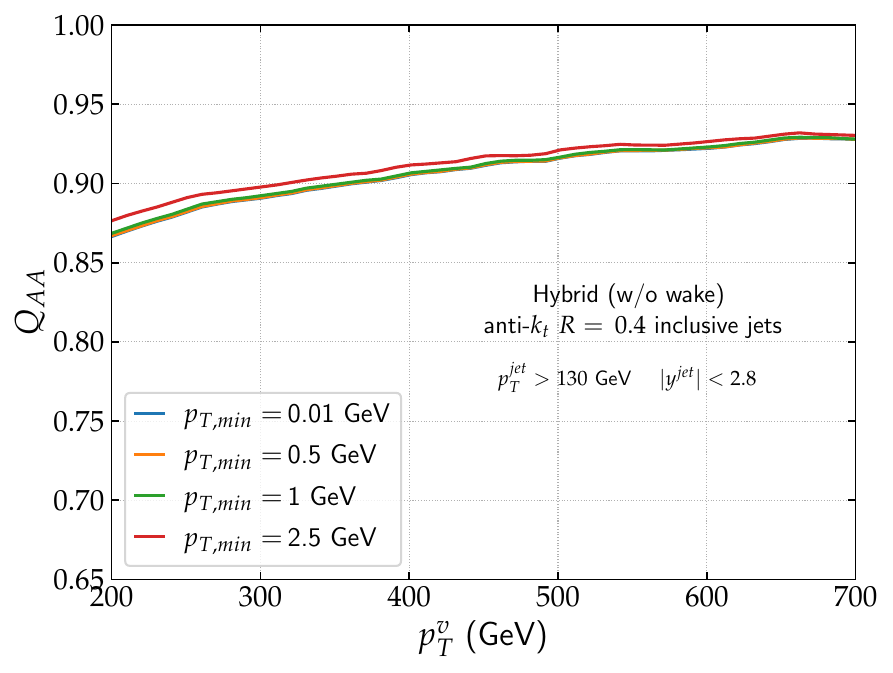}
         \label{fig:three sin x}
     \end{subfigure}
     \caption{$Q_{AA}$ sensitivity to different minimum required $p_T$ of final state particles for {\jewel\ 2.3} (\textbf{left}) and for the Hybrid Model (\textbf{right}). Vacuum samples are pp collisions.}
     \label{fig:ptmin_sens}
\end{figure}

Heavy-ion jets have a larger soft component then their pp counterparts, since both medium response and parton energy loss contribute to an increase of soft particles within the reconstruction cone. As a result (Fig.~\ref{fig:ptmin_sens}), when including medium response, increasing the minimum particle $p_T$ cut increases the energy lost by jets. This effect is very small for $R=0.4$ jets in the Hybrid model, owing to the large angular spread of the wake. Naturally, the increase is greater the larger the jet radius. This is because a larger jet will in principle be more populated in the softest part of the particle spectrum, thus suffering a larger relative energy cut. This radius dependence is much more pronounced in \jewel\ than in the Hybrid Model, which is consistent with what we saw in Section~\ref{subsec:jet_radius}. As was seen in \cite{Casalderrey-Solana:2016jvj}, the implementation of the medium response in the Hybrid Model underestimates the yield of semi-hard particles, leading to an excess of soft particles. This explains why, especially for $R=0.8$, we have a jump from the curve corresponding to a minimum $p_T$ of $0.5$ GeV to the one corresponding to $1$ GeV, which seems to be the range where the $p_T$ of most medium response particles lies. 

\begin{figure}
     \centering
     \begin{subfigure}[h]{0.49\textwidth}
         \centering
         \includegraphics[width=\textwidth]{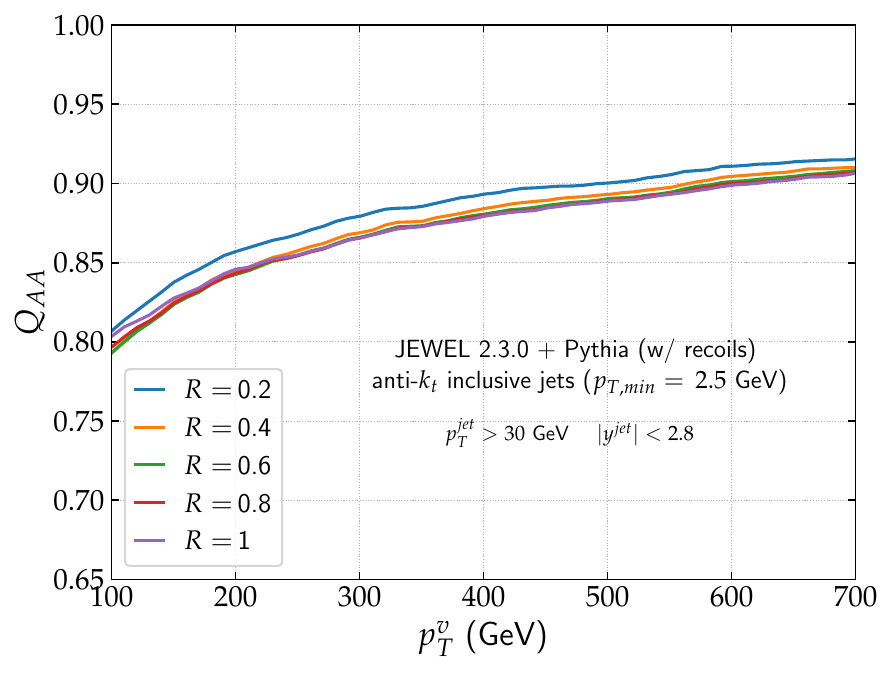}
         \label{fig:three sin x}
     \end{subfigure}
     \begin{subfigure}[h]{0.49\textwidth}
         \centering
         \includegraphics[width=\textwidth]{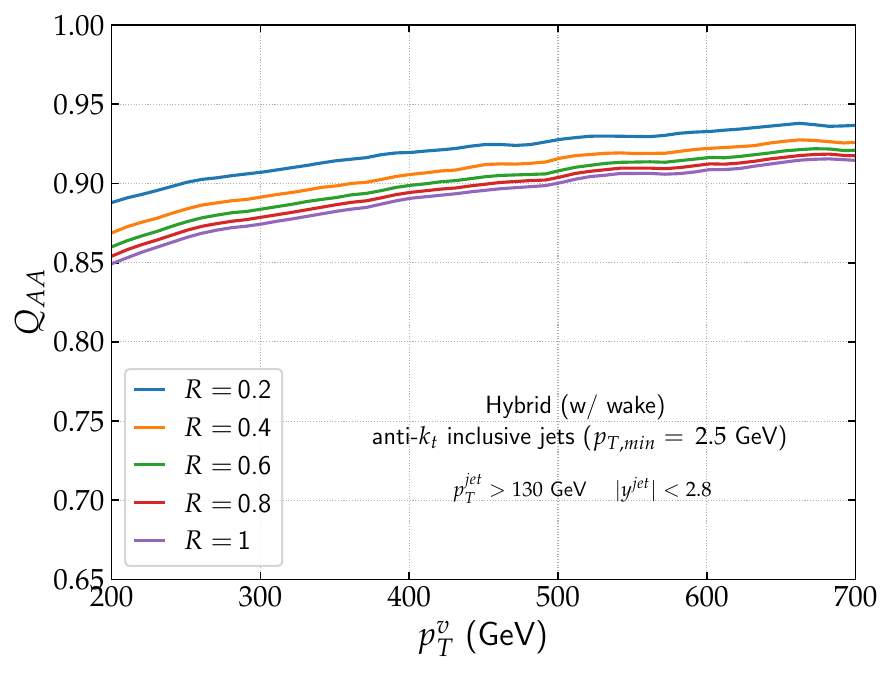}
         \label{fig:three sin x}
     \end{subfigure}
     \caption{Evolution of $Q_{AA}$ jet radius for \jewel\ 2.3 (\textbf{left}) and for the Hybrid Model (\textbf{right}), for an extreme value of minimum required $p_T$ of final state particles $p_{T,min} = 2.5$ GeV). Medium response is included in both models. Vacuum samples are pp collisions.}
     \label{fig:radius_ptmin_sens}
\end{figure}

In the absence of medium response, the sensitivity to minimum $p_T$ of the particles is almost non existent unless one goes to very large values of this cut, as can be seen for the curve corresponding to $p_T = 2.5 $ GeV, which corresponds to a smaller energy loss than the remaining values for minimum $p_T$. Naturally, this arises from the particle constituents of a jet having a harder spectrum in vacuum than in medium. Interestingly enough, at this large a value of minimum $p_T$ and when including medium response, the jet radius dependence of $Q_{AA}$ (Fig.~\ref{fig:radius_ptmin_sens}) is significantly reduced for both models, being almost washed out in \jewel\ (left) except for the smaller jet radii $R\lesssim0.4$. In this case, the Hybrid Model (right) predicts a $Q_{AA}$ with a radius dependence which resembles what we saw for jets in the absence of medium response in the top right panel of Fig.~\ref{fig:qaa_model}. This is expected, since at such a large cut most of the contribution from medium response is discarded. 


~\\

\subsection{Colour charge}\label{sec:color_carge}

We now address the dependence of jet energy loss on the colour charge of the parton, quark or gluon, that initiates a jet. Both perturbative radiative \cite{Baier:1996kr, Baier:1998yf, Baier:1998kq, Zakharov:1996fv, Zakharov:1997uu, Gyulassy:2000fs, Gyulassy:2000er, Wang:2001ifa, Majumder:2009zu, Arnold:2001ms,Arnold:2002ja} and collisonal \cite{Wicks:2005gt} partonic energy losses are proportional to the Casimir colour factor of the propagating parton. This implies that the energy loss of a gluon is larger, by a factor $C_A / C_F = 9/4$, than that of a quark. In the absence of QGP, the amount of energy beyond a given jet reconstruction radius has been shown to not follow Casimir scaling \cite{Dasgupta:2007wa,Dasgupta:2014yra} since the dependence on the Casimir colour factor of the initiating parton is diluted as the parton cascade, dominated by gluon radiation, develops. When QGP is present, gluon initiated jets lose more energy than quark initiated jets by a factor substantially smaller than the $C_A/C_F$ Casimir scaling \cite{Apolinario:2020nyw, Brewer:2020och}.

To establish whether the large observed difference between the $\gamma+$jet $R_{AA}$, where quark jets are dominant, and the more gluon rich  inclusive jet $R_{AA}$ can be explained by the colour charge of the parton initiating a jet, one cannot ignore the existence of $p_T$ bin migration exacerbated by sensitivity of $R_{AA}$ to the steeply falling nature of the jet $p_T$ spectrum. The $\gamma+$jet and inclusive jets $R_{AA}$ could be different not because of different energy loss of quark and gluon jets, but simply because the steepnesses of the jet $p_T$ spectra are different.

\begin{figure}
     \centering
     \begin{subfigure}[h]{0.49\textwidth}
         \centering
         \includegraphics[width=\textwidth]{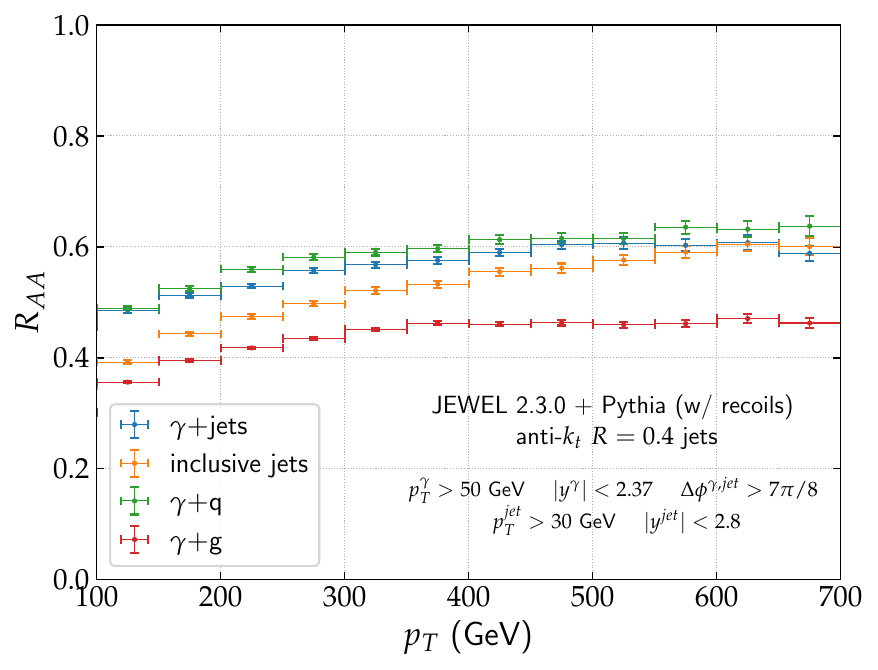}
         \label{dfd}
     \end{subfigure}
     \begin{subfigure}[h]{0.49\textwidth}
         \centering
         \includegraphics[width=\textwidth]{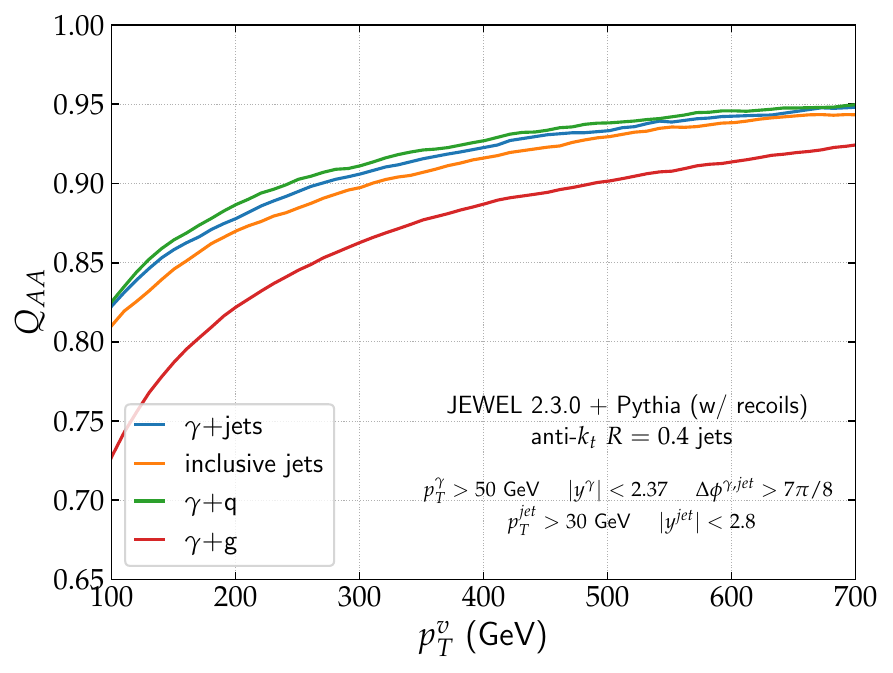}
         \label{fig:three sin x}
     \end{subfigure}
     \caption{Dependence of $R_{AA}$ (\textbf{left}) and $Q_{AA}$ (\textbf{right}) on the hard process originating the jets. Vacuum samples include nuclear effects (isospin averaged nuclear PDFs).}
     \label{fig:process_comp}
\end{figure}

$Q_{AA}$ deconvolutes the effects of $p_T$ migration and spectral shape influence from jet energy loss. In Fig.~\ref{fig:process_comp} we show both $R_{AA}$ and $Q_{AA}$ for $\gamma+$jets, inclusive jets and pure samples of quark jets ($\gamma+$q) and gluon jets ($\gamma+$q). These two last samples were obtained by selecting, at generation level, only the relevant (quark or gluon) channel. The difference between $\gamma+$jets and inclusive jets in $Q_{AA}$ is significantly smaller than for $R_{AA}$ indicating that a significant part of the observed effect in $R_{AA}$ is due to $p_T$ bin migration. Interestingly, the difference in energy loss of pure samples of quark and gluon initiated jets ($\gamma+$q and $\gamma+$g) is partly washed away when comparing $\gamma+$jets and inclusive jets. In fact, inclusive jet $Q_{AA}$ is much closer to quark initiated jet $Q_{AA}$ than to gluon initiated jet $Q_{AA}$ for all $p_T$. While inclusive jets reconstructed in pp collisions are, for  $p_T \lesssim 200 $ GeV, a mixture of quark and gluon initiated jets with fractions of roughly $40/60\,\%$, the larger energy loss of gluon jets in AA leads to a (survivor) bias that leads to a sample of reconstructed jets increasingly dominated by quark jets in each bin of $p_T$. The large steepness of the $p_T$ spectrum further exaggerates this effect. The success of simple parametric models~\cite{Spousta:2015fca, Ogrodnik:2024qug} that encode such a variation of the q/g fraction supports this observation. Nevertheless, it remains to be precisely determined the degree to which $q/g$ fractions in AA collisions are modified with respect to pp on both the theory~\cite{Ringer:2019rfk, Brewer:2020och,Pablos:2022mrx, Zhang:2023oid} and the experimental sides ~\cite{ALICE:2019ykw, CMS:2020plq}. Hence, our results in Fig.~\ref{fig:process_comp} show one or both of the following statements to be true: in-medium $q/g$ fractions are significantly modified due to energy loss; if inside a sample of jets one can distinguish two sub-populations of jets with known fractions and with different energy losses, then the resulting $Q_{AA}$ will differ from the weighted sum of the individual $Q_{AA}$ of each sub-population.

The large difference between inclusive jets and $\gamma+g$ $Q_{AA}$ is in apparent contradiction with what is seen for $R_{AA}$ on the left-hand side plot of Fig.~\ref{fig:process_comp}, where for $p_T \lesssim 200$ GeV the two results are fairly similar. Nevertheless, note that the difference in spectrum steepness and the difference in energy loss affect the hierarchy of $R_{AA}$ in opposite directions. On the one hand, one would expect purely from the spectrum shape, while fixing the same energy loss, that inclusive $R_{AA}$ would be smaller than that of $\gamma+g$, based on the larger steepness of the former. Note that this difference in steepness is dependent on the initial partonic states that go into the leading order matrix element and not only on the final state, jet initiating partons. This means the shape of the inclusive spectrum does not necessarily follow a weighted sum of the $\gamma+q$ and $\gamma+g$ shapes according to the estimated q/g fractions. The energy loss of $\gamma+g$, on the other hand, is greater than that of inclusive jets as per the rough estimation of $40/60\,\%$ for q/g fraction. This would result in a larger $R_{AA}$ for inclusive jets than for purely gluon initiated jets. Hence, what we see for $p_T \lesssim 200$ GeV is a cancellation of these two effects. When calculating $Q_{AA}$ the energy loss is the dominant effect, so the two curves are no longer similar at $p_T \lesssim 200 $ GeV. At large enough $p_T$ the steepness values of both spectra are similar, so the difference in $R_{AA}$ and $Q_{AA}$ are in agreement.

\begin{figure}
     \centering
     \begin{subfigure}[h]{0.49\textwidth}
         \centering
         \includegraphics[width=\textwidth]{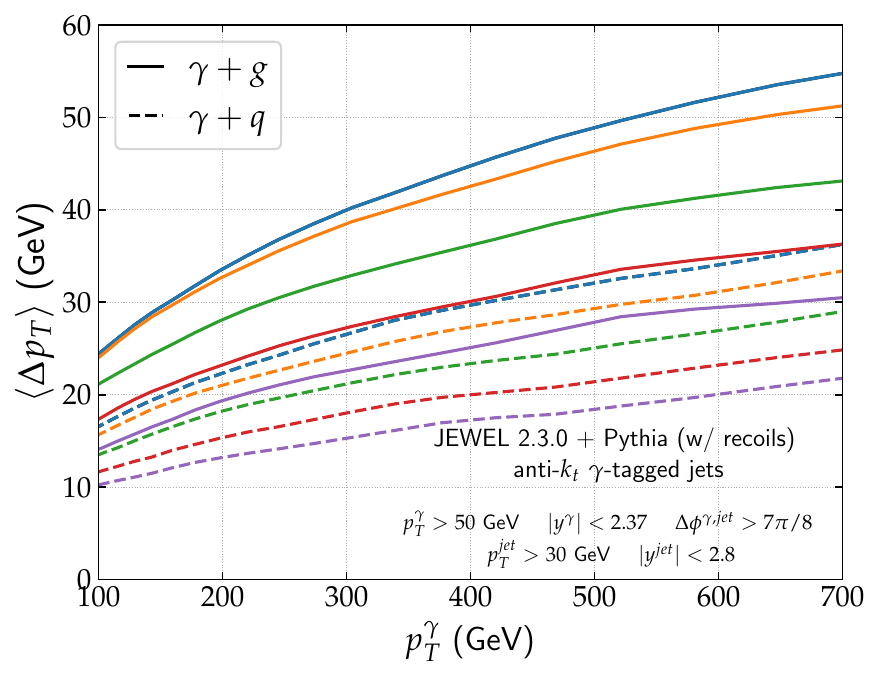}
         \label{dfd}
     \end{subfigure}
     \begin{subfigure}[h]{0.49\textwidth}
         \centering
         \includegraphics[width=\textwidth]{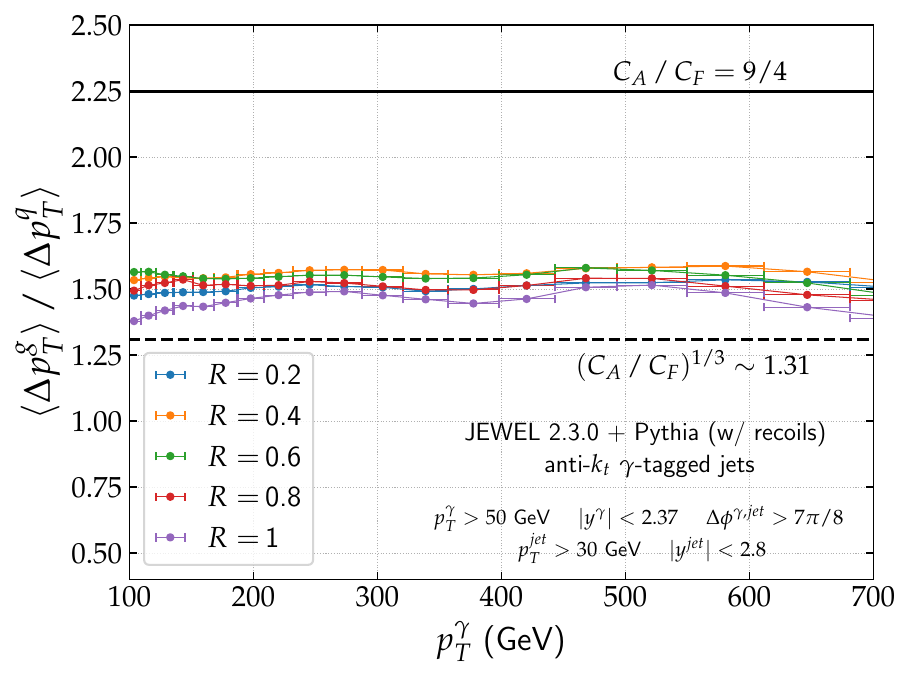}
         \label{fig:three sin x}
     \end{subfigure}
     \caption{Evolution with jet radius of $\langle \Delta p_T \rangle = \langle p_T^v \rangle - \langle p_T^q \rangle $ (see Eqs.~\eqref{eq:av_ptv} and \eqref{eq:av_ptq}) (\textbf{left}) and of the ratio between average energy losses of gluon and quark initiated jets (\textbf{right}), for each bin of $p_T^{\gamma}$. The black horizontal lines corresponds to what is expected from single parton Casimir scaling (solid) and from holography based calculations (dashed) \cite{Casalderrey-Solana:2014bpa, Chesler:2014jva, Chesler:2015nqz}. Vacuum samples include nuclear effects (isospin averaged nuclear PDFs).}
     \label{fig:qg_comp}
\end{figure}
For a more precise statement of the dependence of jet energy loss on the colour charge of the initiating parton, we show on the left-hand side panel of Fig.~\ref{fig:qg_comp} the average absolute energy loss calculated using the quantile procedure for gluon jets (solid lines) and for quark jets (dashed lines), for a given value of photon $p_T$.
Making the energy loss a function of photon $p_T$ ensures that we avoid the bias coming from comparing quark and gluon jets with the different initial $p_T$ in vacuum. The first interesting observation is that the jet radius dependence of energy loss is much more pronounced in gluon jets than in quarks jets. This is to be expected as per previous arguments, since a gluon radiating more means gluon initiated jets have a larger angular spread. Then, it is clear that gluon jets lose more energy than quarks jets for a wide range of jet radii. 
The ratio between gluon and quark jet energy loss for different radii is shown on right-hand side panel of Fig.~\ref{fig:qg_comp}. It is stable around $\sim 1.5$ independently of both jet radius and jet $p_T$.
The Casimir scaling of single partons (the black line at $9/4$) is substantially diluted in jets, with gluon jets not losing much more energy than quarks jets.
Interestingly, the holographic calculations of single parton energy loss \cite{Casalderrey-Solana:2014bpa, Chesler:2014jva, Chesler:2015nqz} that underlie the Hybrid Model predict a single parton scaling which is much milder than Casimir scaling ($\sim 1.31$) and which is much closer to our result obtained for in-medium jets. Note that this same result could be obtained by considering the energy loss proxy in Eq.~\eqref{eq:av_ptv} obtained simply by using the photon $p_T$ to match mean jet $p_T$ in vacuum and medium rather than the one obtained through quantile matching. Indeed, although we don't show it here, this provides another testament to the quantile procedure's robustness, since the result obtained using Eq.~\eqref{eq:av_ptv} shows the same features as the right-hand side plot of Fig~\ref{fig:qg_comp}.

\section{Towards a $Q_{AA}$ measurement}\label{sec:measurement}

An obvious experimental obstacle to the calculation of the quantile function as defined in Eq.~\eqref{eq:quantile_matching} is the requirement to integrate jet spectra up to arbitrarily large $p_T$, which is naturally not possible in experiment. However, doing the integration with a cutoff at $+\infty$ is not actually a requirement, since one can cast Eq.~\eqref{eq:quantile_matching} into a form that takes into account $p_T$ cutoffs in both spectra. This is done by considering Eq.~\eqref{eq:quantile_matching} at two different values of $p_T^v$ and subtracting the two resulting equations, giving $\Sigma^{pp}(p_{T,1}^v, p_{T,2}^v) = \Sigma^{AA}(p_T^{q}(p_{T,1}^v),p_T^{q}(p_{T,2}^v))$.
Hence, if the pp spectrum has a cutoff at a value $p_{T,2}^v=p_{T,c}^v$ then the upper integration limit of $\Sigma^{AA}$ should be $p_{T,c}^q = p_T^{q}(p_{T,c}^v)$:
\begin{equation}\label{eq:ptc_m}
    \int_{p_{T}^v}^{p_{T,c}^v} dp_T \,  \frac{d\sigma^{pp}}{dp_T} = \int_{p_T^q(p_{T}^v)}^{p_{T,c}^q} dp_T \,  \frac{d\sigma^{AA}}{dp_T}\, . 
\end{equation}
However, because the right-hand side depends on the unknown quantile function evaluated at two different points ($p_{T,c}^v$ and $p_T^v$), the equation is underdetermined. This is expected, since Eq.~\eqref{eq:ptc_m} is an integral equation and therefore needs an initial condition that sets the value for $p_{T,c}^q$. For values of $p_T^v$ much smaller than $p_{T,c}^v$, the sensitivity to the upper integration limit for the medium spectrum is insignificant and one can go around this problem by setting $p_{T,c}^q = p_{T,c}^v$. For values comparable to $p_T^c$, however, the value of this limit becomes relevant. This is clearly seen in the left plot of Fig.~\ref{fig:cutoff_sensitivity}, where we see that setting the wrong value for $p_{T,c}^q$ (blue curve -- overestimate it by imposing same cutoff; orange curve -- underestimate it by setting it to $90\%$ of the correct cutoff $p_{T,c}^q$) results in a significant reduction in $Q_{AA}$ accuracy for a range of about $150$ GeV below the vacuum cutoff.

\subsection{Cutoff dependence}\label{sec:cutoff}
\begin{figure}
    \centering
     \begin{subfigure}[h]{0.49\textwidth}
         \centering
         \includegraphics[width=\textwidth]{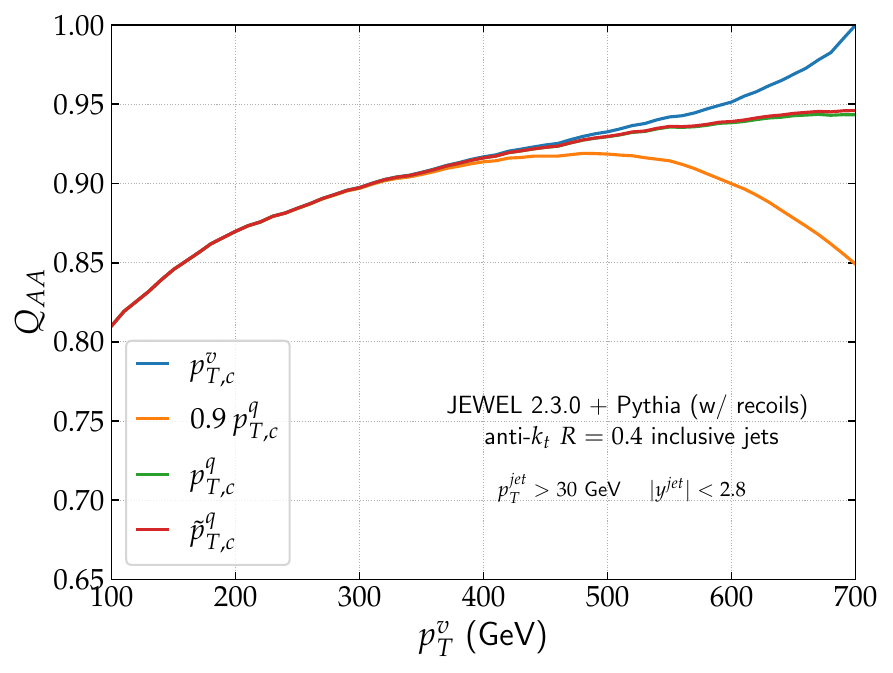}
         \label{dfd}
     \end{subfigure}
     \begin{subfigure}[h]{0.49\textwidth}
         \centering
         \includegraphics[width=\textwidth]{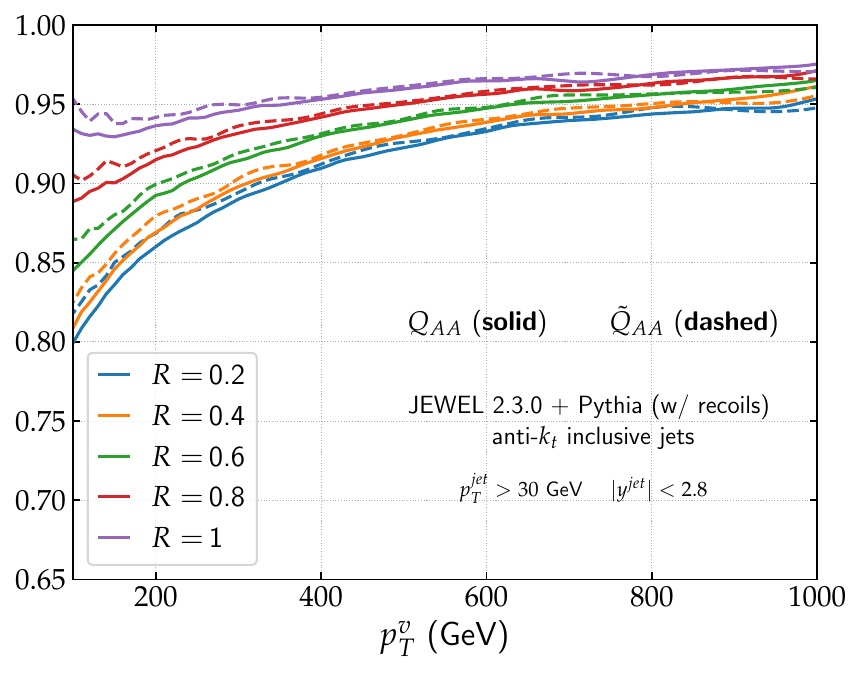}
         \label{fig:three sin x}
     \end{subfigure}
    \caption{\textbf{Left: }$Q_{AA}$ sensitivity to different values of $p_{T,c}^m$ in Eq.~\eqref{eq:ptc_m}. \textbf{Right: }Comparison of $Q_{AA}$ and $\tilde Q_{AA}$ for multiple jet radii. Vacuum samples include nuclear effects (isospin averaged nuclear PDFs).}
    \label{fig:cutoff_sensitivity}
\end{figure}

Possible workarounds for this problem include extracting $p_T^q(p_{T,c}^v)$ from an interpolated quantile function calculated via Monte Carlo at larger $p_T$ than what is experimentally accessible. Another option, which can be fully calculated from data, is to use a variant of the quantile function also introduced in the original paper \cite{Brewer:2018dfs}, the pseudo-quantile $\tilde p_T^q$, to get an estimate of the initial condition $p_{T,c}^q = p_T^q(p_T^c)$. The way to calculate the pseudo-quantile is to equate the spectra themselves instead of their cumulative versions:
\begin{equation}\label{eq:pqaa_def}
    \frac{d\sigma^{AA}}{d\tilde p_T^q}(\tilde{p}_T^{q}(p_T^v))= \frac{d\sigma^{pp}}{dp_T^v}(p_T^v) \, .
\end{equation}
Interestingly enough, under the same assumptions as for the quantile function $p_T^q$ in Eq.~\eqref{eq:qaa_estimate}, one gets:
\begin{equation}\label{eq:pqaa_estimate}
  1 - \frac{\tilde p_T^q(p_T^v)}{p_T^v} = 1-\tilde{Q}_{AA} (p_T^v) \approx \frac{\langle \epsilon \rangle (p_T^v)}{p_T^v} \, .
\end{equation}
Hence, despite having a clear conceptual difference, the quantile and the pseudo-quantile functions are in good agreement under the assumption of small and only mildly $p_T$-dependent energy loss. Indeed, by looking at Eq.~\eqref{eq:diff_qaa}, we see that it is equivalent to Eq.~\eqref{eq:pqaa_def} in the limit where we have some constant energy loss shift $\epsilon_0$
\begin{equation}
p_T^q(p_T^v) \rightarrow p_T^v + \epsilon_0\, .
\end{equation}
It is then expected that when calculating $\tilde Q_{AA}$ and comparing it to $Q_{AA}$ (right-hand side plot in Fig.~\ref{fig:cutoff_sensitivity}), the two are not strikingly different for most of the $p_T$ range and that they are in great agreement at large enough $p_T$ and for all jet radii. By looking at the red curve of the left-hand side plot of Fig.~\ref{fig:cutoff_sensitivity}, we then confirm that using $\tilde p_{T,c}^q = \tilde p_T^q (p_T^c)$ as an estimate for the initial condition $ p_{T,c}^q = p_T^q(p_{T,c}^v)$ works well.

\subsection{Equivalence between $1-Q_{AA}$ and $S_{loss}$}
\label{subsec:sloss_qaa}
As discussed at the end of Section~\ref{sec:qaa_intro}, a relation similar to our Eq.~\eqref{eq:diff_qaa} was used in \cite{ATLAS:2023iad} to estimate jet energy loss. Renaming the variables, it reads
\begin{align}\label{eq:atlas_eq}
    \frac{d\sigma^{AA}}{dp_T^s}(p_T^s) = \frac{d\sigma^{pp}}{dp_T^v}(p_T^v)\left(1+\frac{d \Delta p_T^s}{dp_T^v}\right)\, ,
\end{align}
where $p_T^s(p_T^v) = p_T^v-\Delta p_T^s (p_T^v) = (1-S_{loss}(p_T^v))p_T^v$. Setting $\Delta p_T^q(p_T^v) \equiv p_T^v-p_T^q(p_T^v) = (1-Q_{AA}(p_T^v)p_T^v$, the differential equation defining the quantile matching in Eq.~\eqref{eq:diff_qaa} can be re-written as
\begin{align}\label{eq:diff_qaa_rewritten}
    \frac{d\sigma^{AA}}{dp_T^q}(p_T^q) = \frac{d\sigma^{pp}}{dp_T^v}(p_T^v)\left(1-\frac{d \Delta p_T^q}{dp_T^v}\right)^{-1} \, .
\end{align}

\begin{figure}
    \centering
     \begin{subfigure}[h]{0.49\textwidth}
         \centering
         \includegraphics[width=\textwidth]{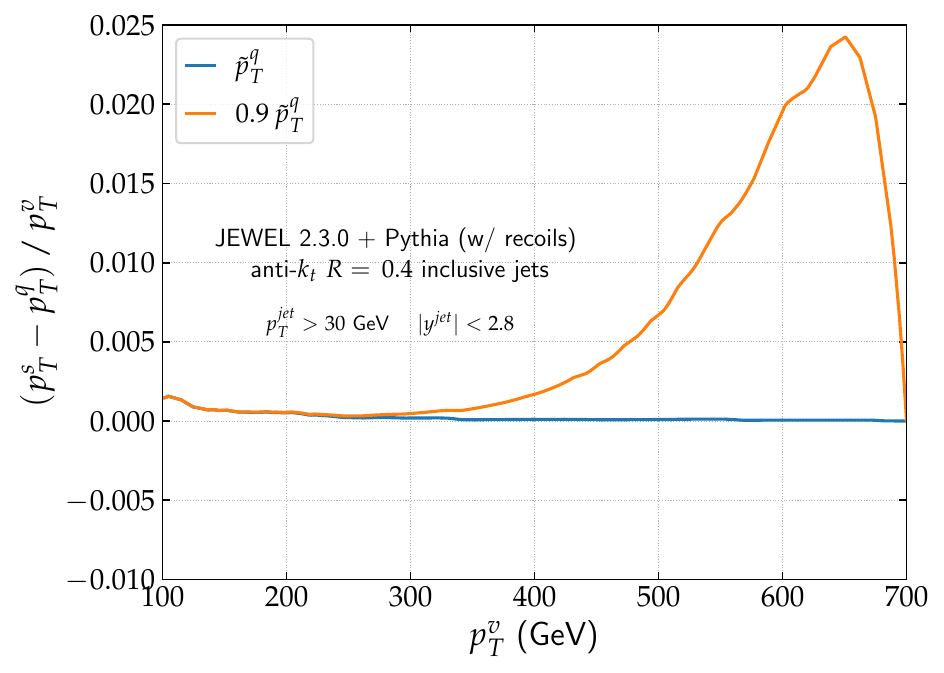}
         \label{dfd}
     \end{subfigure}
     \begin{subfigure}[h]{0.49\textwidth}
         \centering
         \includegraphics[width=\textwidth]{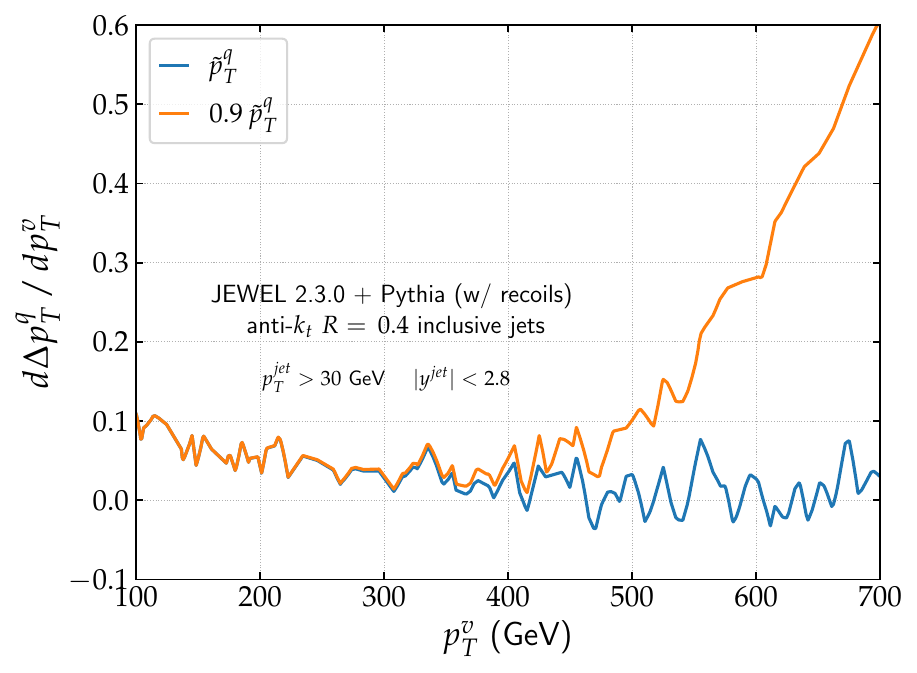}
         \label{fig:three sin x}
     \end{subfigure}
    \caption{\textbf{Left: }Comparing energy loss proxies $\Delta p_T^q = p_T^v - p_T^q$ in Eq.~\eqref{eq:diff_qaa_rewritten} and $\Delta p_T^s = p_T^v - p_T^s$ in Eq.~\eqref{eq:atlas_eq} for two differential values of initial condition given by $\tilde p_T^q (p_T^v = 700 \text{ GeV})$ (reasonable estimate) and by $0.9\, \tilde p_T^q (p_T^v=700 \text{ GeV})$ (underestimate). \textbf{Right: } Derivative of the energy loss function calculated through Eq.~\eqref{eq:diff_qaa_rewritten} for the same two values of initial condition. Vacuum samples include nuclear effects (isospin averaged nuclear PDFs).}
    \label{fig:proxy_comp}
\end{figure}

Looking at Eqs.~\eqref{eq:atlas_eq} and \eqref{eq:diff_qaa_rewritten}, we see that both agree to first order in $d\Delta p_T/dp_T^v$, thus computing similar proxies ($p_T^s \approx p_T^q$) for a mildly $p_T$-dependent energy loss. Evidently, both quantities agree with the pseudo-quantile as defined in Eq.~\eqref{eq:pqaa_def} to zeroth order in $d\Delta p_T/dp_T^v$ (see right plot of Fig. \ref{fig:cutoff_sensitivity}). Nevertheless, if energy loss has a stronger $p_T$ dependence than in  {\jewel\ 2.3}, then the two equations may give significantly different results. Additionally, this dependence can be artificially introduced if one incorrectly sets the initial condition. This is clearly seen in Fig.~\ref{fig:proxy_comp} (left plot), where we show the difference between the two energy loss proxies $(\Delta p_T^q - \Delta p_T^s) \, / \, p_T^v  = (p_T^s - p_T^q) \, / \, p_T^v$. For the blue curve we estimate the initial condition using the pseudo-quantile as defined in Eq.~\eqref{eq:pqaa_def}, which we have seen provides an accurate approximation to the quantile at large enough $p_T$ (see Fig.~\ref{fig:cutoff_sensitivity}). In this case, Eqs.~\eqref{eq:atlas_eq} and \eqref{eq:diff_qaa_rewritten} calculate almost exactly the same quantity, with the largest difference ($\sim 0.1\%$ of the lost energy) being at lowest end of the $p_T$ range ($\sim 100$ GeV). This is expected since, as can be seen on the right plot of Fig~\ref{fig:proxy_comp}, the derivative appearing on the right hand side of both equations is quite small -- it is always below $0.1$ and decreases with increasing $p_T^v$. If we provide an inaccurate estimate for the initial condition, e.g., $0.9$ of the previous value, then the two quantities $\Delta p_T^s$ and $\Delta p_T^q$ separate from each other, with the largest difference reaching $\sim 2.5\%$ of the lost energy. Again, we can explain this by looking at the right plot of Fig.~\ref{fig:proxy_comp}, where we see the derivative increasing past a value for which Eqs.~\eqref{eq:diff_qaa_rewritten} and \eqref{eq:atlas_eq} agree to first order. Despite these differences, it is clear the quantities $1-Q_{AA}$ and $S_{loss}$ should not be too different from each other when applied to experimental data and when reasonably estimating the initial condition. Nevertheless, the quantile procedure has a much clearer physical motivation and, although not explored here, it provides more solid groundwork for the study of jet energy loss dispersion.

~\\

\section{Jet substructure observables after quantile correction}\label{sec:substructure}
\begin{figure}
     \centering
     \begin{subfigure}[h]{0.49\textwidth}
         \centering
         \includegraphics[width=\textwidth]{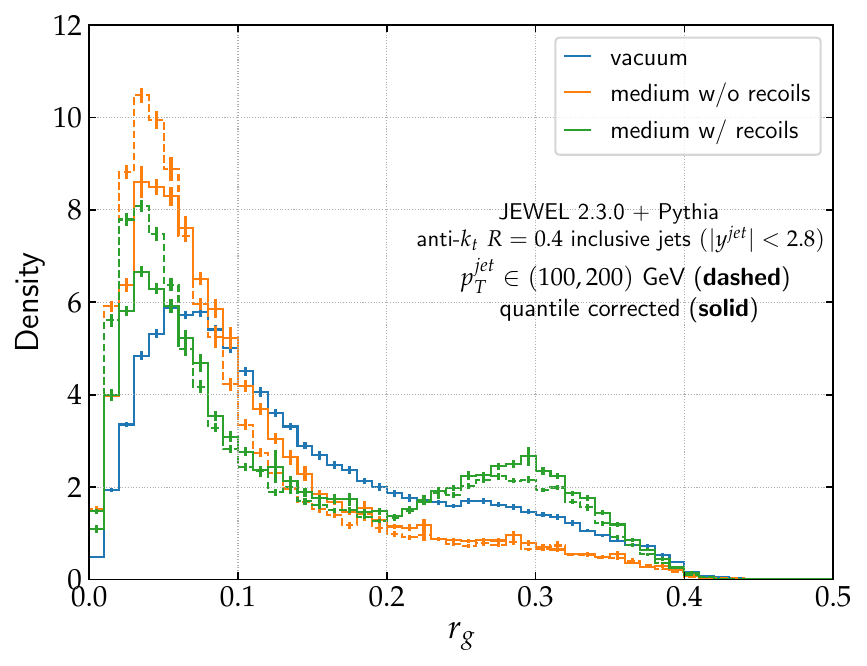}
         \label{dfd}
     \end{subfigure}
     \begin{subfigure}[h]{0.49\textwidth}
         \centering
         \includegraphics[width=\textwidth]{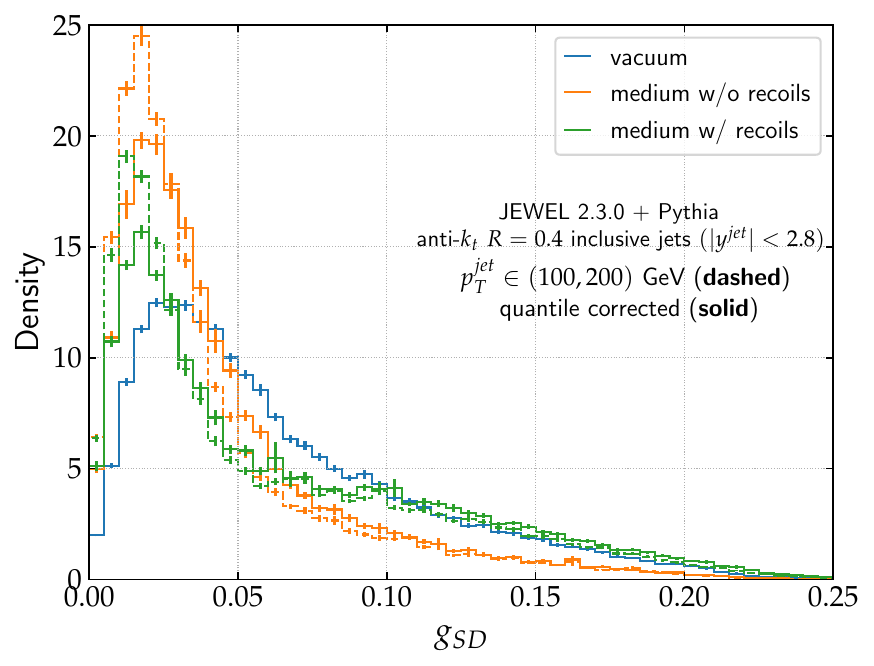}
         \label{fig:three sin x}
     \end{subfigure}
     \begin{subfigure}[h]{0.49\textwidth}
         \centering
         \includegraphics[width=\textwidth]{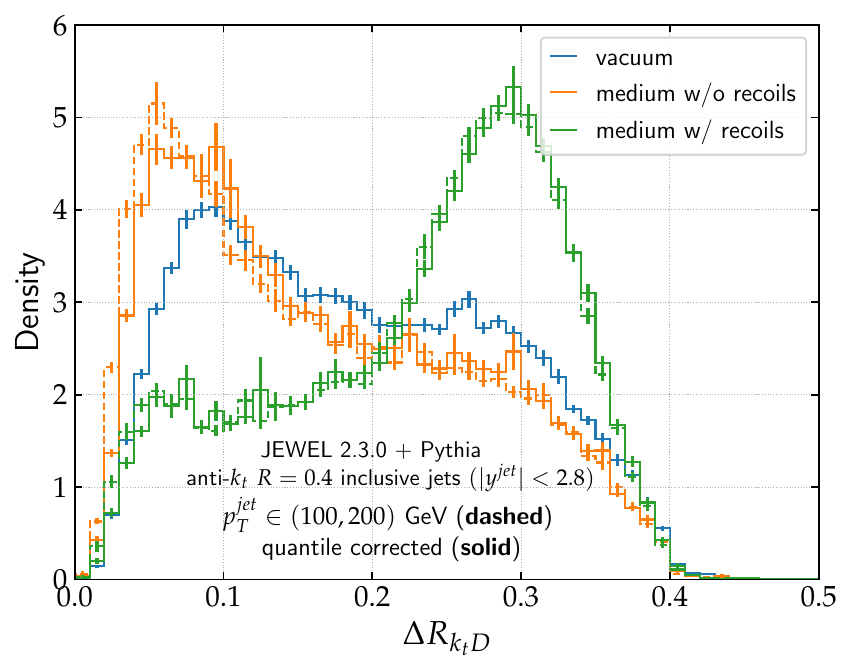}
         \label{dfd}
     \end{subfigure}
     \begin{subfigure}[h]{0.49\textwidth}
         \centering
         \includegraphics[width=\textwidth]{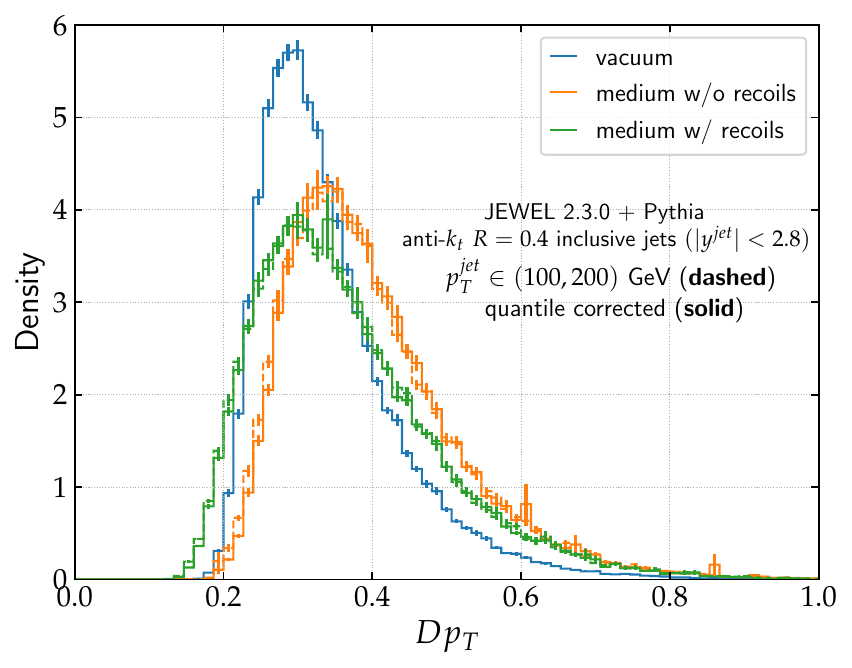}
         \label{fig:three sin x}
     \end{subfigure}
     \caption{Distribution of four jet substructure observables for vacuum, medium without recoils and medium with recoils. The solid lines correspond to jets with $p_T^{jet} \in (100, 200)$ GeV and the dashed lines to $p_T^q \in (79,170)$ GeV (medium w/o recoils) and $p_T^q \in (84,177)$ GeV (medium w/ recoils). Vacuum samples are pp collisions.}
     \label{fig:corrected_obs}
\end{figure}

The main purpose of the quantile matching procedure is to suppress the $p_T$ migration bias associated with observable comparison within the same $p_T^{jet}$ bin. By estimating average jet energy loss, one compares vacuum and medium jets in matched $p_T$ bins thus comparing jets that were, on average, initiated by hard partons with the same $p_T$. Naturally, studying how sensitive to $p_T$ migration bias an observable is depends on the correlation of such observable with jet $p_T$. This result was showed in the original paper \cite{Brewer:2018dfs}, where the authors saw the impact of $p_T$ migration on the ratio between jet mass and jet transverse momentum $m^{jet}/p_{T}^{jet}$. Let us extend this analysis to other jet observables and also include medium response as implemented in \jewel. In Fig.~\ref{fig:corrected_obs}, we show the distribution of four jet substructure observables for jets in a fixed $p_T$ bin (dashed lines) and for quantile matched $p_T$ bins (solid lines). The list of interesting jet substructure observables is quite extensive \cite{Andrews:2018jcm}, but we decided to chose a few which illustrate both observables whose distributions are significantly corrected and those which are practically insensitive to the $p_T$ migration.

Let us first focus on the groomed jet radius $r_g$ (top left corner in Fig.~\ref{fig:corrected_obs}). To calculate $r_g$, one first reclusters a jet using the Cambridge-Aachen (C/A) algorithm and then calculates the angular distance $\Delta R_{ij}$ of the first branching/subjet that satisfies the SoftDrop condition \cite{Larkoski:2014wba}
\begin{equation}\label{eq:SD_condition}
    \frac{\text{min} [p_{T,i},p_{T,j}]}{p_{T,i} + p_{T,j}} > z_{cut}\left(\frac{\Delta R_{ij}}{R}\right)^{\beta}\, ,
\end{equation}
where $z_{cut}=0.1$ and $\beta = 0$. We see that for $r_g$ the contribution of $p_T$ migration is quite significant both with and without medium response. In fact, the apparent narrowing of jets from vacuum to medium is reduced when correcting the $p_T$ bin with quantile matching. Continuing with an observable which conveys information about the width of a jet, we now look at the jet girth $g_{SD}$ (top right corner in Fig.~\ref{fig:corrected_obs}) after applying SoftDrop:
\begin{equation}
    g_{SD} = \sum_{i \in jet_{SD}} z_i \,\Delta R_{i,jet}\, ,
\end{equation}
where $z_i$ is the fraction of transverse momentum of the constituent $i$ with respect to jet $p_T$ and $\Delta R_{i,jet}$ is the angular distance of the constituent with respect to the jet axis. Applying SoftDrop to groom the jet amounts to discarding the softest branch at each step of the C/A sequence until the condition in Eq.~\eqref{eq:SD_condition} is met. Looking at the result, the solid curves (quantile matched bins) are significantly deviated from the dashed ones (unmatched bins), such that the apparent narrowing of jets is, like for $r_g$, reduced. This is true whether or not we include medium recoils. 

As for observables that are not as sensitive to $p_T$ migration (i.e. that are only weakly correlated with $p_T$ in the $p_T$ range we consider), let us look at yet another observable quantifying the angular spread of a jet -- the distance $\Delta R$ between the two subjets in the C/A reclustering sequence that satisfy the following dynamical grooming \cite{Mehtar-Tani:2019rrk} condition:
\begin{equation}
    \kappa^{(a)} = \frac{1}{p_{T}^{jet}} \mathop{\text{max}}_{i \in \text{C/A seq.}} \left[z_i(1-z_i)p_{T,i}\left(\frac{\theta_i}{R}\right)^a\right] \, .
\end{equation}
By setting $a = 1$ ($k_T$Drop) we are tagging the branching with the largest relative transverse momentum $k_T \sim \kappa^{(1)}\, p_T^{jet}$. Looking at bottom left corner of Fig.~\ref{fig:corrected_obs}, we see that in the absence of medium response this observable is only slightly sensitive to $p_T$ migration (orange curves). However, when including medium response (green curves) the $p_T$ correlation seems to be washed out since the observable is very little modified after quantile matching the $p_T$ bin. Finally, in lower right corner of Fig.~\ref{fig:corrected_obs}, we show the momentum dispersion inside a jet, $Dp_T$, which quantifies how hard or soft the jet fragmentation is:
\begin{equation}
    Dp_{T} = \frac{\sqrt{\mathop{\sum}_{i\in jet} p_{T,i}^2}}{p_{T}^{jet}} \, .
\end{equation}
Whether or not we include medium recoils, this observable is almost insensitive to $p_T$ migration.

\section{Conclusions}\label{sec:conclusions}
Performing an unbiased comparison of proton-proton and heavy-ion jet populations is essential to be able to infer quantitative knowledge about the transport properties of the underlying QGP. In this work, and following the groundwork laid by \cite{Brewer:2018dfs}, we took a step in this direction by further exploring how the quantile procedure can be used as a tool to mitigate the $p_T$ migration bias inherent to usual jet comparisons by providing a proxy for average fractional jet energy loss ($Q_{AA}$). 

Comparing $Q_{AA}$ with the jet energy loss proxy obtained by using the photon's $p_T$ as a reference in $\gamma+$jet events, we concluded that the agreement between the two is reasonable, even in the presence of a large source of fluctuations, the medium response, especially pronounced for larger jet radii. One should note, however, that the former systematically underestimates energy loss compared to the latter. The evolution of $Q_{AA}$ with jet radius was also studied for two different models of jet evolution in heavy-ion collisions -- \jewel\ and the Hybrid Model. In the absence of medium response, the two models exhibit the same behaviour of $Q_{AA}$ with jet radius -- larger jets lose more energy --, indicating, for \jewel, that the increase in jet components contributing to energy loss dominates over the recapture of medium-induced radiation inside the jet cone. When including medium response, however, the difference in implementations appears quite significant since the two models predict quite different behaviours with jet radius. Overall, the conclusion is that medium jets in \jewel\ recapture energy at a significantly larger, and quite possibly exaggerated, rate than in the Hybrid Model, since in the latter the medium response is not only softer but also more widely distributed in angle. For similar reasons, jets in \jewel\ exhibit a $Q_{AA}$ which is much more sensitive to minimum particle $p_T$ than in the Hybrid Model. By studying this sensitivity at a larger jet radius, we clearly see that the $p_T$ of particles in the medium response of the Hybrid Model is dominated by the range $\sim(0.5,1)$ GeV. With an extreme cut at $2.5$ GeV, the medium response seems to be removed in the Hybrid model, while this is not the case in \jewel.

The strength of colour charge dependence was also evaluated by comparing the $R_{AA}$ and $Q_{AA}$ of $\gamma+$jets and inclusive jets. This lead to the conclusion that spectrum shape severely impacts this comparison when looking at $R_{AA}$, since disentangling this effect from the actual energy loss using $Q_{AA}$ reveals that colour charge dependence is not as strong as predicted by the large observed difference in jet suppression. By further comparing these results with the $Q_{AA}$ of pure samples of quark jets ($\gamma+$q) and of gluon jets ($\gamma+$q), we conclude that one or both of the following statements must be true: the in-medium $q/g$ fractions are modified with respect to the vacuum due to the survivor bias induced by gluon jets losing more energy than quarks jets; the $Q_{AA}$ of a population of jets containing two sub-populations with substantially different energy losses is not trivially given by the weighted sum of their individual energy losses. Additionally, we compared the ratio between the absolute energy loss of gluon jets and that of quarks jets with the expected Casimir scaling of single parton energy loss. The result is substantially smaller than the Casimir ratio $C_F/C_A$ and it closer to the value predicted by holographic calculations of parton energy loss (e.g. ~\cite{Casalderrey-Solana:2014bpa, Chesler:2015nqz}).

The experimental obstacle posed by the requirement to integrate the jet $p_T$ spectrum up to $+\infty$ to calculate $Q_{AA}$ was addressed. In fact, this requirement can be translated into the need for an initial condition of the equation defining the quantile function, which can be quite accurately estimated, for large enough $p_T$, using the pseudo-quantile function~\cite{Brewer:2018dfs}, which uses the differential cross-sections directly in the computation. A comparison between $Q_{AA}$ and $S_{loss}$, which was used in~\cite{ATLAS:2023iad}, as jet energy loss proxies was drawn. Put simply, we find no advantage in using 
$S_{loss}$ rather than $Q_{AA}$ to estimate jet energy loss, given that the two observables involve the same technical difficulties (namely providing an accurate initial condition) and $Q_{AA}$ has a clear physical motivation from which one can build more complex, rigorous energy loss studies.
At the end, a simple exploration of the size of $p_T$ migration in a few substructure observables was carried out, illustrating the potential of the quantile procedure to quantify jet modifications due to interaction with a QGP in a more unbiased way.

As a last note, we point out a clear limitation of  $Q_{AA}$ as currently defined. 
It is certainly tempting to calculate $Q_{AA}$ in bins of a given jet substructure observable, that we will refer to as $X$. In doing so, we would be going beyond just an average value of energy loss, partially resolving energy loss dispersion by correlating it with a single jet substructure observable. However, doing so would effectively get us into the same type of bias that exists when comparing jets in bins of equal reconstructed $p_T$. In particular, this would mean that we would be confounding energy loss of jets with the modification of the distribution of $X$ in the medium with respect to the vacuum. A possible solution would be to calculate a quantile matching in $X$, assuming its bin migration to be monotonic. Nevertheless, bin migration in $X$ and in $p_T$ are not, in principle, independent. Hence, to calculate them separately would be to distort this dependence, obtaining results which do not reflect actual energy loss dependence on substructure. Having said this, the only rigorously correct procedure to take is to calculate a quantile matching in $p_T$ and $X$ simultaneously. However, this problem is undetermined since we would have a single (quantile) equation for two unknown (quantile) functions. In order for this to be possible, further assumptions about the nature of bin migration in $X$ and its dependence on $p_T$ would need to be made.

\acknowledgments

We thank Daniel Pablos and Arjun Srinivasan Kudinoor for kindly assisting us with the Hybrid model samples and for discussions. This work is supported by European Research Council project ERC-2018-ADG-835105 YoctoLHC. It as also been supported by OE - Portugal, Fundação para a Ciência e a Tecnologia (FCT), I.P., under projects CERN/FIS-PAR/0032/2021 (http://doi.org/10.54499/CERN/FIS-PAR/0032/2021) and EXPL/FIS-PAR/0905/2021 (https://doi.org/10.54499/EXPL/FIS-PAR/0905/2021). L.A. was supported by FCT under contract 2021.03209.CEECIND. L.L. was supported by grant id 46225 from EXPL/FIS-PAR/0905/2021. J.M.S. was supported by OE - Portugal, Fundação para a Ciência e Tecnologia (FCT) under contract PRT/BD/152262/2021.

\newpage
\appendix
\bibliographystyle{jhep}
\bibliography{main.bib}

\providecommand{\href}[2]{#2}\begingroup\raggedright\begin{thebibliography}{10}

\bibitem{PHENIX:2001hpc}
{\bf PHENIX} Collaboration, K.~Adcox et~al., {\it {Suppression of hadrons with
  large transverse momentum in central Au+Au collisions at $\sqrt{s_{NN}}$ =
  130-GeV}},  {\em Phys. Rev. Lett.} {\bf 88} (2002) 022301,
  [\href{http://arxiv.org/abs/nucl-ex/0109003}{{\tt nucl-ex/0109003}}].

\bibitem{STAR:2005gfr}
{\bf STAR} Collaboration, J.~Adams et~al., {\it {Experimental and theoretical
  challenges in the search for the quark gluon plasma: The STAR Collaboration's
  critical assessment of the evidence from RHIC collisions}},  {\em Nucl. Phys.
  A} {\bf 757} (2005) 102--183,
  [\href{http://arxiv.org/abs/nucl-ex/0501009}{{\tt nucl-ex/0501009}}].

\bibitem{STAR:2020xiv}
{\bf STAR} Collaboration, J.~Adam et~al., {\it {Measurement of inclusive
  charged-particle jet production in Au + Au collisions at $\sqrt{s_{NN}}=$200
  GeV}},  {\em Phys. Rev. C} {\bf 102} (2020), no.~5 054913,
  [\href{http://arxiv.org/abs/2006.00582}{{\tt arXiv:2006.00582}}].

\bibitem{Muller:2012zq}
B.~Muller, J.~Schukraft, and B.~Wyslouch, {\it {First Results from Pb+Pb
  collisions at the LHC}},  {\em Ann. Rev. Nucl. Part. Sci.} {\bf 62} (2012)
  361--386, [\href{http://arxiv.org/abs/1202.3233}{{\tt arXiv:1202.3233}}].

\bibitem{ATLAS:2018gwx}
{\bf ATLAS} Collaboration, M.~Aaboud et~al., {\it {Measurement of the nuclear
  modification factor for inclusive jets in Pb+Pb collisions at
  $\sqrt{s_\mathrm{NN}}=5.02$ TeV with the ATLAS detector}},  {\em Phys. Lett.
  B} {\bf 790} (2019) 108--128, [\href{http://arxiv.org/abs/1805.05635}{{\tt
  arXiv:1805.05635}}].

\bibitem{CMS:2021vui}
{\bf CMS} Collaboration, A.~M. Sirunyan et~al., {\it {First measurement of
  large area jet transverse momentum spectra in heavy-ion collisions}},  {\em
  JHEP} {\bf 05} (2021) 284, [\href{http://arxiv.org/abs/2102.13080}{{\tt
  arXiv:2102.13080}}].

\bibitem{Busza:2018rrf}
W.~Busza, K.~Rajagopal, and W.~van~der Schee, {\it {Heavy Ion Collisions: The
  Big Picture, and the Big Questions}},  {\em Ann. Rev. Nucl. Part. Sci.} {\bf
  68} (2018) 339--376, [\href{http://arxiv.org/abs/1802.04801}{{\tt
  arXiv:1802.04801}}].

\bibitem{Apolinario:2022vzg}
L.~Apolin\'ario, Y.-J. Lee, and M.~Winn, {\it {Heavy quarks and jets as probes
  of the QGP}},  {\em Prog. Part. Nucl. Phys.} {\bf 127} (2022) 103990,
  [\href{http://arxiv.org/abs/2203.16352}{{\tt arXiv:2203.16352}}].

\bibitem{Zapp:2012ak}
K.~C. Zapp, F.~Krauss, and U.~A. Wiedemann, {\it {A perturbative framework for
  jet quenching}},  {\em JHEP} {\bf 03} (2013) 080,
  [\href{http://arxiv.org/abs/1212.1599}{{\tt arXiv:1212.1599}}].

\bibitem{Caucal:2018dla}
P.~Caucal, E.~Iancu, A.~H. Mueller, and G.~Soyez, {\it {Vacuum-like jet
  fragmentation in a dense QCD medium}},  {\em Phys. Rev. Lett.} {\bf 120}
  (2018) 232001, [\href{http://arxiv.org/abs/1801.09703}{{\tt
  arXiv:1801.09703}}].

\bibitem{Mehtar-Tani:2013pia}
Y.~Mehtar-Tani, J.~G. Milhano, and K.~Tywoniuk, {\it {Jet physics in heavy-ion
  collisions}},  {\em Int. J. Mod. Phys. A} {\bf 28} (2013) 1340013,
  [\href{http://arxiv.org/abs/1302.2579}{{\tt arXiv:1302.2579}}].

\bibitem{Qin:2015srf}
G.-Y. Qin and X.-N. Wang, {\it {Jet quenching in high-energy heavy-ion
  collisions}},  {\em Int. J. Mod. Phys. E} {\bf 24} (2015), no.~11 1530014,
  [\href{http://arxiv.org/abs/1511.00790}{{\tt arXiv:1511.00790}}].

\bibitem{Cunqueiro:2021wls}
L.~Cunqueiro and A.~M. Sickles, {\it {Studying the QGP with Jets at the LHC and
  RHIC}},  {\em Prog. Part. Nucl. Phys.} {\bf 124} (2022) 103940,
  [\href{http://arxiv.org/abs/2110.14490}{{\tt arXiv:2110.14490}}].

\bibitem{Apolinario:2024equ}
L.~Apolin\'ario, Y.-T. Chien, and L.~Cunqueiro~Mendez, {\it {Jet
  substructure}},  {\em Int. J. Mod. Phys. E} {\bf 33} (2024), no.~07 2430003.

\bibitem{Wicks:2005gt}
S.~Wicks, W.~Horowitz, M.~Djordjevic, and M.~Gyulassy, {\it {Elastic,
  inelastic, and path length fluctuations in jet tomography}},  {\em Nucl.
  Phys. A} {\bf 784} (2007) 426--442,
  [\href{http://arxiv.org/abs/nucl-th/0512076}{{\tt nucl-th/0512076}}].

\bibitem{Gyulassy:2000fs}
M.~Gyulassy, P.~Levai, and I.~Vitev, {\it {NonAbelian energy loss at finite
  opacity}},  {\em Phys. Rev. Lett.} {\bf 85} (2000) 5535--5538,
  [\href{http://arxiv.org/abs/nucl-th/0005032}{{\tt nucl-th/0005032}}].

\bibitem{Gyulassy:2000er}
M.~Gyulassy, P.~Levai, and I.~Vitev, {\it {Reaction operator approach to
  nonAbelian energy loss}},  {\em Nucl. Phys. B} {\bf 594} (2001) 371--419,
  [\href{http://arxiv.org/abs/nucl-th/0006010}{{\tt nucl-th/0006010}}].

\bibitem{Baier:1996kr}
R.~Baier, Y.~L. Dokshitzer, A.~H. Mueller, S.~Peigne, and D.~Schiff, {\it
  {Radiative energy loss of high-energy quarks and gluons in a finite volume
  quark - gluon plasma}},  {\em Nucl. Phys. B} {\bf 483} (1997) 291--320,
  [\href{http://arxiv.org/abs/hep-ph/9607355}{{\tt hep-ph/9607355}}].

\bibitem{Baier:1998yf}
R.~Baier, Y.~L. Dokshitzer, A.~H. Mueller, and D.~Schiff, {\it {Radiative
  energy loss of high-energy partons traversing an expanding QCD plasma}},
  {\em Phys. Rev. C} {\bf 58} (1998) 1706--1713,
  [\href{http://arxiv.org/abs/hep-ph/9803473}{{\tt hep-ph/9803473}}].

\bibitem{Baier:1998kq}
R.~Baier, Y.~L. Dokshitzer, A.~H. Mueller, and D.~Schiff, {\it {Medium induced
  radiative energy loss: Equivalence between the BDMPS and Zakharov
  formalisms}},  {\em Nucl. Phys. B} {\bf 531} (1998) 403--425,
  [\href{http://arxiv.org/abs/hep-ph/9804212}{{\tt hep-ph/9804212}}].

\bibitem{Zakharov:1996fv}
B.~G. Zakharov, {\it {Fully quantum treatment of the Landau-Pomeranchuk-Migdal
  effect in QED and QCD}},  {\em JETP Lett.} {\bf 63} (1996) 952--957,
  [\href{http://arxiv.org/abs/hep-ph/9607440}{{\tt hep-ph/9607440}}].

\bibitem{Zakharov:1997uu}
B.~G. Zakharov, {\it {Radiative energy loss of high-energy quarks in finite
  size nuclear matter and quark - gluon plasma}},  {\em JETP Lett.} {\bf 65}
  (1997) 615--620, [\href{http://arxiv.org/abs/hep-ph/9704255}{{\tt
  hep-ph/9704255}}].

\bibitem{Wang:2001ifa}
X.-N. Wang and X.-f. Guo, {\it {Multiple parton scattering in nuclei: Parton
  energy loss}},  {\em Nucl. Phys. A} {\bf 696} (2001) 788--832,
  [\href{http://arxiv.org/abs/hep-ph/0102230}{{\tt hep-ph/0102230}}].

\bibitem{Majumder:2009zu}
A.~Majumder, {\it {The In-medium scale evolution in jet modification}},
  \href{http://arxiv.org/abs/0901.4516}{{\tt arXiv:0901.4516}}.

\bibitem{Arnold:2001ms}
P.~B. Arnold, G.~D. Moore, and L.~G. Yaffe, {\it {Photon emission from quark
  gluon plasma: Complete leading order results}},  {\em JHEP} {\bf 12} (2001)
  009, [\href{http://arxiv.org/abs/hep-ph/0111107}{{\tt hep-ph/0111107}}].

\bibitem{Arnold:2002ja}
P.~B. Arnold, G.~D. Moore, and L.~G. Yaffe, {\it {Photon and gluon emission in
  relativistic plasmas}},  {\em JHEP} {\bf 06} (2002) 030,
  [\href{http://arxiv.org/abs/hep-ph/0204343}{{\tt hep-ph/0204343}}].

\bibitem{Mehtar-Tani:2010ebp}
Y.~Mehtar-Tani, C.~A. Salgado, and K.~Tywoniuk, {\it {Anti-angular ordering of
  gluon radiation in QCD media}},  {\em Phys. Rev. Lett.} {\bf 106} (2011)
  122002, [\href{http://arxiv.org/abs/1009.2965}{{\tt arXiv:1009.2965}}].

\bibitem{Mehtar-Tani:2021fud}
Y.~Mehtar-Tani, D.~Pablos, and K.~Tywoniuk, {\it {Cone-Size Dependence of Jet
  Suppression in Heavy-Ion Collisions}},  {\em Phys. Rev. Lett.} {\bf 127}
  (2021), no.~25 252301, [\href{http://arxiv.org/abs/2101.01742}{{\tt
  arXiv:2101.01742}}].

\bibitem{CDF:2005prv}
{\bf CDF} Collaboration, D.~Acosta et~al., {\it {Study of jet shapes in
  inclusive jet production in $p\bar{p}$ collisions at $\sqrt{s}=1.96$ TeV}},
  {\em Phys. Rev. D} {\bf 71} (2005) 112002,
  [\href{http://arxiv.org/abs/hep-ex/0505013}{{\tt hep-ex/0505013}}].

\bibitem{He:2018gks}
Y.~He, L.-G. Pang, and X.-N. Wang, {\it {Bayesian extraction of jet energy loss
  distributions in heavy-ion collisions}},  {\em Phys. Rev. Lett.} {\bf 122}
  (2019), no.~25 252302, [\href{http://arxiv.org/abs/1808.05310}{{\tt
  arXiv:1808.05310}}].

\bibitem{Wu:2023azi}
J.~Wu, W.~Ke, and X.-N. Wang, {\it {Bayesian inference of the path-length
  dependence of jet energy loss}},  {\em Phys. Rev. C} {\bf 108} (2023), no.~3
  034911, [\href{http://arxiv.org/abs/2304.06339}{{\tt arXiv:2304.06339}}].

\bibitem{Zhang:2023oid}
S.-L. Zhang, E.~Wang, H.~Xing, and B.-W. Zhang, {\it {Flavor dependence of jet
  quenching in heavy-ion collisions from a Bayesian analysis}},  {\em Phys.
  Lett. B} {\bf 850} (2024) 138549,
  [\href{http://arxiv.org/abs/2303.14881}{{\tt arXiv:2303.14881}}].

\bibitem{Connors:2017ptx}
M.~Connors, C.~Nattrass, R.~Reed, and S.~Salur, {\it {Jet measurements in heavy
  ion physics}},  {\em Rev. Mod. Phys.} {\bf 90} (2018) 025005,
  [\href{http://arxiv.org/abs/1705.01974}{{\tt arXiv:1705.01974}}].

\bibitem{Brewer:2021hmh}
J.~Brewer, Q.~Brodsky, and K.~Rajagopal, {\it {Disentangling jet modification
  in jet simulations and in Z+jet data}},  {\em JHEP} {\bf 02} (2022) 175,
  [\href{http://arxiv.org/abs/2110.13159}{{\tt arXiv:2110.13159}}].

\bibitem{Andres:2024hdd}
C.~Andres, J.~Holguin, R.~Kunnawalkam~Elayavalli, and J.~Viinikainen, {\it
  {Minimizing Selection Bias in Inclusive Jets in Heavy-Ion Collisions with
  Energy Correlators}},  \href{http://arxiv.org/abs/2409.07514}{{\tt
  arXiv:2409.07514}}.

\bibitem{Brewer:2018dfs}
J.~Brewer, J.~G. Milhano, and J.~Thaler, {\it {Sorting out quenched jets}},
  {\em Phys. Rev. Lett.} {\bf 122} (2019), no.~22 222301,
  [\href{http://arxiv.org/abs/1812.05111}{{\tt arXiv:1812.05111}}].

\bibitem{ATLAS:2023iad}
{\bf ATLAS} Collaboration, G.~Aad et~al., {\it {Comparison of inclusive and
  photon-tagged jet suppression in 5.02 TeV Pb+Pb collisions with ATLAS}},
  {\em Phys. Lett. B} {\bf 846} (2023) 138154,
  [\href{http://arxiv.org/abs/2303.10090}{{\tt arXiv:2303.10090}}].

\bibitem{Milhano:2015mng}
J.~G. Milhano and K.~C. Zapp, {\it {Origins of the di-jet asymmetry in heavy
  ion collisions}},  {\em Eur. Phys. J. C} {\bf 76} (2016), no.~5 288,
  [\href{http://arxiv.org/abs/1512.08107}{{\tt arXiv:1512.08107}}].

\bibitem{Rajagopal:2016uip}
K.~Rajagopal, A.~V. Sadofyev, and W.~van~der Schee, {\it {Evolution of the jet
  opening angle distribution in holographic plasma}},  {\em Phys. Rev. Lett.}
  {\bf 116} (2016), no.~21 211603, [\href{http://arxiv.org/abs/1602.04187}{{\tt
  arXiv:1602.04187}}].

\bibitem{Brewer:2017fqy}
J.~Brewer, K.~Rajagopal, A.~Sadofyev, and W.~Van Der~Schee, {\it {Evolution of
  the Mean Jet Shape and Dijet Asymmetry Distribution of an Ensemble of
  Holographic Jets in Strongly Coupled Plasma}},  {\em JHEP} {\bf 02} (2018)
  015, [\href{http://arxiv.org/abs/1710.03237}{{\tt arXiv:1710.03237}}].

\bibitem{Casalderrey-Solana:2016jvj}
J.~Casalderrey-Solana, D.~Gulhan, G.~Milhano, D.~Pablos, and K.~Rajagopal, {\it
  {Angular Structure of Jet Quenching Within a Hybrid Strong/Weak Coupling
  Model}},  {\em JHEP} {\bf 03} (2017) 135,
  [\href{http://arxiv.org/abs/1609.05842}{{\tt arXiv:1609.05842}}].

\bibitem{Takacs:2021bpv}
A.~Takacs and K.~Tywoniuk, {\it {Quenching effects in the cumulative jet
  spectrum}},  {\em JHEP} {\bf 10} (2021) 038,
  [\href{http://arxiv.org/abs/2103.14676}{{\tt arXiv:2103.14676}}].

\bibitem{Baier:2001yt}
R.~Baier, Y.~L. Dokshitzer, A.~H. Mueller, and D.~Schiff, {\it {Quenching of
  hadron spectra in media}},  {\em JHEP} {\bf 09} (2001) 033,
  [\href{http://arxiv.org/abs/hep-ph/0106347}{{\tt hep-ph/0106347}}].

\bibitem{PHENIX:2004vcz}
{\bf PHENIX} Collaboration, K.~Adcox et~al., {\it {Formation of dense partonic
  matter in relativistic nucleus-nucleus collisions at RHIC: Experimental
  evaluation by the PHENIX collaboration}},  {\em Nucl. Phys. A} {\bf 757}
  (2005) 184--283, [\href{http://arxiv.org/abs/nucl-ex/0410003}{{\tt
  nucl-ex/0410003}}].

\bibitem{Zapp:2013vla}
K.~C. Zapp, {\it {JEWEL 2.0.0: directions for use}},  {\em Eur. Phys. J. C}
  {\bf 74} (2014), no.~2 2762, [\href{http://arxiv.org/abs/1311.0048}{{\tt
  arXiv:1311.0048}}].

\bibitem{Bossi:2024qho}
H.~Bossi, A.~S. Kudinoor, I.~Moult, D.~Pablos, A.~Rai, and K.~Rajagopal, {\it
  {Imaging the Wakes of Jets with Energy-Energy-Energy Correlators}},
  \href{http://arxiv.org/abs/2407.13818}{{\tt arXiv:2407.13818}}.

\bibitem{Casalderrey-Solana:2014bpa}
J.~Casalderrey-Solana, D.~C. Gulhan, J.~G. Milhano, D.~Pablos, and
  K.~Rajagopal, {\it {A Hybrid Strong/Weak Coupling Approach to Jet
  Quenching}},  {\em JHEP} {\bf 10} (2014) 019,
  [\href{http://arxiv.org/abs/1405.3864}{{\tt arXiv:1405.3864}}]. [Erratum:
  JHEP 09, 175 (2015)].

\bibitem{Casalderrey-Solana:2015vaa}
J.~Casalderrey-Solana, D.~C. Gulhan, J.~G. Milhano, D.~Pablos, and
  K.~Rajagopal, {\it {Predictions for Boson-Jet Observables and Fragmentation
  Function Ratios from a Hybrid Strong/Weak Coupling Model for Jet Quenching}},
   {\em JHEP} {\bf 03} (2016) 053, [\href{http://arxiv.org/abs/1508.00815}{{\tt
  arXiv:1508.00815}}].

\bibitem{Hulcher:2017cpt}
Z.~Hulcher, D.~Pablos, and K.~Rajagopal, {\it {Resolution Effects in the Hybrid
  Strong/Weak Coupling Model}},  {\em JHEP} {\bf 03} (2018) 010,
  [\href{http://arxiv.org/abs/1707.05245}{{\tt arXiv:1707.05245}}].

\bibitem{Casalderrey-Solana:2018wrw}
J.~Casalderrey-Solana, Z.~Hulcher, G.~Milhano, D.~Pablos, and K.~Rajagopal,
  {\it {Simultaneous description of hadron and jet suppression in heavy-ion
  collisions}},  {\em Phys. Rev. C} {\bf 99} (2019), no.~5 051901,
  [\href{http://arxiv.org/abs/1808.07386}{{\tt arXiv:1808.07386}}].

\bibitem{Milhano:2022kzx}
J.~G. Milhano and K.~Zapp, {\it {Improved background subtraction and a fresh
  look at jet sub-structure in JEWEL}},  {\em Eur. Phys. J. C} {\bf 82} (2022),
  no.~11 1010, [\href{http://arxiv.org/abs/2207.14814}{{\tt
  arXiv:2207.14814}}].

\bibitem{Sjostrand:2014zea}
T.~Sj\"ostrand, S.~Ask, J.~R. Christiansen, R.~Corke, N.~Desai, P.~Ilten,
  S.~Mrenna, S.~Prestel, C.~O. Rasmussen, and P.~Z. Skands, {\it {An
  introduction to PYTHIA 8.2}},  {\em Comput. Phys. Commun.} {\bf 191} (2015)
  159--177, [\href{http://arxiv.org/abs/1410.3012}{{\tt arXiv:1410.3012}}].

\bibitem{Chesler:2014jva}
P.~M. Chesler and K.~Rajagopal, {\it {Jet quenching in strongly coupled
  plasma}},  {\em Phys. Rev. D} {\bf 90} (2014), no.~2 025033,
  [\href{http://arxiv.org/abs/1402.6756}{{\tt arXiv:1402.6756}}].

\bibitem{Chesler:2015nqz}
P.~M. Chesler and K.~Rajagopal, {\it {On the Evolution of Jet Energy and
  Opening Angle in Strongly Coupled Plasma}},  {\em JHEP} {\bf 05} (2016) 098,
  [\href{http://arxiv.org/abs/1511.07567}{{\tt arXiv:1511.07567}}].

\bibitem{PhysRevD.10.186}
F.~Cooper and G.~Frye, {\it Single-particle distribution in the hydrodynamic
  and statistical thermodynamic models of multiparticle production},  {\em
  Phys. Rev. D} {\bf 10} (Jul, 1974) 186--189.

\bibitem{nPDFs}
K.~J. Eskola, P.~Paakkinen, H.~Paukkunen, and C.~A. Salgado, {\it {EPPS16:
  Nuclear parton distributions with LHC data}},  {\em Eur. Phys. J. C} {\bf 77}
  (2017), no.~3 163, [\href{http://arxiv.org/abs/1612.05741}{{\tt
  arXiv:1612.05741}}].

\bibitem{PDFs}
S.~Dulat, T.-J. Hou, J.~Gao, M.~Guzzi, J.~Huston, P.~Nadolsky, J.~Pumplin,
  C.~Schmidt, D.~Stump, and C.~P. Yuan, {\it {New parton distribution functions
  from a global analysis of quantum chromodynamics}},  {\em Phys. Rev. D} {\bf
  93} (2016), no.~3 033006, [\href{http://arxiv.org/abs/1506.07443}{{\tt
  arXiv:1506.07443}}].

\bibitem{lhapdf}
A.~Buckley, J.~Ferrando, S.~Lloyd, K.~Nordstr\"om, B.~Page, M.~R\"ufenacht,
  M.~Sch\"onherr, and G.~Watt, {\it {LHAPDF6: parton density access in the LHC
  precision era}},  {\em Eur. Phys. J. C} {\bf 75} (2015) 132,
  [\href{http://arxiv.org/abs/1412.7420}{{\tt arXiv:1412.7420}}].

\bibitem{Sjostrand:2006za}
T.~Sjostrand, S.~Mrenna, and P.~Z. Skands, {\it {PYTHIA 6.4 Physics and
  Manual}},  {\em JHEP} {\bf 05} (2006) 026,
  [\href{http://arxiv.org/abs/hep-ph/0603175}{{\tt hep-ph/0603175}}].

\bibitem{Li:2019dre}
H.~T. Li and I.~Vitev, {\it {Jet charge modification in dense QCD matter}},
  {\em Phys. Rev. D} {\bf 101} (2020) 076020,
  [\href{http://arxiv.org/abs/1908.06979}{{\tt arXiv:1908.06979}}].

\bibitem{Eskola:2009uj}
K.~J. Eskola, H.~Paukkunen, and C.~A. Salgado, {\it {EPS09: A New Generation of
  NLO and LO Nuclear Parton Distribution Functions}},  {\em JHEP} {\bf 04}
  (2009) 065, [\href{http://arxiv.org/abs/0902.4154}{{\tt arXiv:0902.4154}}].

\bibitem{Cacciari:2008gp}
M.~Cacciari, G.~P. Salam, and G.~Soyez, {\it {The anti-$k_t$ jet clustering
  algorithm}},  {\em JHEP} {\bf 04} (2008) 063,
  [\href{http://arxiv.org/abs/0802.1189}{{\tt arXiv:0802.1189}}].

\bibitem{fastjet}
M.~Cacciari, G.~P. Salam, and G.~Soyez, {\it {FastJet User Manual}},  {\em Eur.
  Phys. J. C} {\bf 72} (2012) 1896, [\href{http://arxiv.org/abs/1111.6097}{{\tt
  arXiv:1111.6097}}].

\bibitem{KunnawalkamElayavalli:2017hxo}
R.~Kunnawalkam~Elayavalli and K.~C. Zapp, {\it {Medium response in JEWEL and
  its impact on jet shape observables in heavy ion collisions}},  {\em JHEP}
  {\bf 07} (2017) 141, [\href{http://arxiv.org/abs/1707.01539}{{\tt
  arXiv:1707.01539}}].

\bibitem{Dasgupta:2007wa}
M.~Dasgupta, L.~Magnea, and G.~P. Salam, {\it {Non-perturbative QCD effects in
  jets at hadron colliders}},  {\em JHEP} {\bf 02} (2008) 055,
  [\href{http://arxiv.org/abs/0712.3014}{{\tt arXiv:0712.3014}}].

\bibitem{Dasgupta:2014yra}
M.~Dasgupta, F.~Dreyer, G.~P. Salam, and G.~Soyez, {\it {Small-radius jets to
  all orders in QCD}},  {\em JHEP} {\bf 04} (2015) 039,
  [\href{http://arxiv.org/abs/1411.5182}{{\tt arXiv:1411.5182}}].

\bibitem{Apolinario:2020nyw}
L.~Apolin\'ario, J.~a. Barata, and G.~Milhano, {\it {On the breaking of Casimir
  scaling in jet quenching}},  {\em Eur. Phys. J. C} {\bf 80} (2020), no.~6
  586, [\href{http://arxiv.org/abs/2003.02893}{{\tt arXiv:2003.02893}}].

\bibitem{Brewer:2020och}
J.~Brewer, J.~Thaler, and A.~P. Turner, {\it {Data-driven quark and gluon jet
  modification in heavy-ion collisions}},  {\em Phys. Rev. C} {\bf 103} (2021),
  no.~2 L021901, [\href{http://arxiv.org/abs/2008.08596}{{\tt
  arXiv:2008.08596}}].

\bibitem{Spousta:2015fca}
M.~Spousta and B.~Cole, {\it {Interpreting single jet measurements in Pb $+$ Pb
  collisions at the LHC}},  {\em Eur. Phys. J. C} {\bf 76} (2016), no.~2 50,
  [\href{http://arxiv.org/abs/1504.05169}{{\tt arXiv:1504.05169}}].

\bibitem{Ogrodnik:2024qug}
A.~Ogrodnik, M.~Ryb\'a\v{r}, and M.~Spousta, {\it {Flavor and path-length
  dependence of jet quenching from inclusive jet and \ensuremath{\gamma}-jet
  suppression}},  \href{http://arxiv.org/abs/2407.11234}{{\tt
  arXiv:2407.11234}}.

\bibitem{Ringer:2019rfk}
F.~Ringer, B.-W. Xiao, and F.~Yuan, {\it {Can we observe jet $P_T$-broadening
  in heavy-ion collisions at the LHC?}},  {\em Phys. Lett. B} {\bf 808} (2020)
  135634, [\href{http://arxiv.org/abs/1907.12541}{{\tt arXiv:1907.12541}}].

\bibitem{Pablos:2022mrx}
D.~Pablos and A.~Soto-Ontoso, {\it {Pushing forward jet substructure
  measurements in heavy-ion collisions}},  {\em Phys. Rev. D} {\bf 107} (2023),
  no.~9 094003, [\href{http://arxiv.org/abs/2210.07901}{{\tt
  arXiv:2210.07901}}].

\bibitem{ALICE:2019ykw}
{\bf ALICE} Collaboration, S.~Acharya et~al., {\it {Exploration of jet
  substructure using iterative declustering in pp and Pb\textendash{}Pb
  collisions at LHC energies}},  {\em Phys. Lett. B} {\bf 802} (2020) 135227,
  [\href{http://arxiv.org/abs/1905.02512}{{\tt arXiv:1905.02512}}].

\bibitem{CMS:2020plq}
{\bf CMS} Collaboration, A.~M. Sirunyan et~al., {\it {Measurement of quark- and
  gluon-like jet fractions using jet charge in PbPb and pp collisions at 5.02
  TeV}},  {\em JHEP} {\bf 07} (2020) 115,
  [\href{http://arxiv.org/abs/2004.00602}{{\tt arXiv:2004.00602}}].

\bibitem{Andrews:2018jcm}
H.~A. Andrews et~al., {\it {Novel tools and observables for jet physics in
  heavy-ion collisions}},  {\em J. Phys. G} {\bf 47} (2020), no.~6 065102,
  [\href{http://arxiv.org/abs/1808.03689}{{\tt arXiv:1808.03689}}].

\bibitem{Larkoski:2014wba}
A.~J. Larkoski, S.~Marzani, G.~Soyez, and J.~Thaler, {\it {Soft Drop}},  {\em
  JHEP} {\bf 05} (2014) 146, [\href{http://arxiv.org/abs/1402.2657}{{\tt
  arXiv:1402.2657}}].

\bibitem{Mehtar-Tani:2019rrk}
Y.~Mehtar-Tani, A.~Soto-Ontoso, and K.~Tywoniuk, {\it {Dynamical grooming of
  QCD jets}},  {\em Phys. Rev. D} {\bf 101} (2020), no.~3 034004,
  [\href{http://arxiv.org/abs/1911.00375}{{\tt arXiv:1911.00375}}].

\end{thebibliography}\endgroup

\end{document}